\documentclass{elsarticle}

\usepackage[dvipsnames]{xcolor}
\usepackage[backgroundcolor=magenta!10!white]{todonotes}

\usepackage{microtype}
\usepackage{amsmath,amssymb,amsfonts}

\usepackage{graphicx}

\usepackage{ifthen}

\usepackage{numprint}
\npdecimalsign{.}
\npthousandsep{,}

\usepackage{hyperref}

\usepackage{thm-restate}

\usepackage{tikz}
\usetikzlibrary{arrows,
automata,
calc,
decorations,
decorations.pathreplacing,
positioning,
shapes}

\newtheorem{definition}{Definition}
\newtheorem{theorem}[definition]{Theorem}
\newtheorem{lemma}[definition]{Lemma}
\newtheorem{proposition}[definition]{Proposition}
\newtheorem{corollary}[definition]{Corollary}
\newtheorem{example}[definition]{Example}
\newtheorem{remark}{Remark}[section]

\newproof{proof}{Proof}

\usepackage{caption}
\usepackage{lscape}

\def\<{\langle}
\def\>{\rangle}

\def\Nat{\mathbb{N}}

\def\Real{\mathbb{R}}

\def\cA{\mathcal{A}}
\def\cB{\mathcal{B}}

\def\cD{\mathcal{D}}

\def\cL{\mathcal{L}}
\def\cM{\mathcal{M}}

\def\cU{\mathcal{U}}

\def\Pr{\mathrm{Pr}}

\def\Paths{\mathit{Paths}}

\def\Acc{\mathit{Acc}}

\newcommand{\subpointone}{\,$<\!\psec{0.1}$}

\newcommand{\until}{\cU}
\newcommand{\finally}{\Diamond}
\newcommand{\globally}{\Box}
\newcommand{\neXt}{\bigcirc}

\newcommand{\psec}[1]{\nprounddigits{1}\npfourdigitnosep\numprint[s]{#1}}
\newcommand{\pnodes}[1]{\nprounddigits{0}\numprint{#1}}

\newcommand{\prism}{\texttt{PRISM}}
\newcommand{\colt}{\texttt{COLT}}
\newcommand{\spot}{\texttt{SPOT}}
\newcommand{\ltltodstar}{\texttt{ltl2dstar}}
\newcommand{\ltltotgba}{\texttt{ltl2tgba}}
\newcommand{\rabinizer}{\texttt{Rabinizer}}
\newcommand{\tulip}{\texttt{Tulip}}

\newcommand{\explicitengine}{\texttt{explicit}}

\colorlet{davidColor}{YellowGreen!30!white}
\colorlet{stefanColor}{Brown!30!white}
\colorlet{benColor}{Yellow!30!white}

\newcommand{\vzero}{\vec{0}}%
\newcommand{\vone}{\vec{1}}%
\newcommand{\tB}{\overline{B}}

\newcommand{\Ex}{\mathrm{Ex}}
\newcommand{\then}{\mathord{\triangleright}}

\begin{document}
\begin{abstract}
  Unambiguous automata are nondeterministic automata in which every
  word has at most one accepting run.  In this paper we give a
  polynomial-time algorithm for model checking discrete-time Markov
  chains against $\omega$-regular specifications represented as unambiguous
  automata.  We furthermore show that the complexity of this model
  checking problem lies in NC: the subclass of P comprising
  those problems solvable in poly-logarithmic parallel time.  These
  complexity bounds match the known bounds for model checking Markov
  chains against specifications given as deterministic automata,
  notwithstanding the fact that unambiguous automata can be
  exponentially more succinct than deterministic automata.  We report
  on an implementation of our procedure, including an experiment in
  which the implementation is used to model check LTL formulas on
  Markov chains.
\end{abstract}
\pagestyle{headings}  

\title{Markov Chains and Unambiguous Automata}

\author[tud]{Christel Baier\texorpdfstring{\fnref{tudthanks}}{}}
\author[oxf]{Stefan Kiefer\texorpdfstring{\fnref{kieferthanks}}{}}
\author[tud]{Joachim Klein\texorpdfstring{\fnref{tudthanks}}{}}
\author[tud]{David M\"uller\texorpdfstring{\fnref{tudthanks}}{}}
\author[oxf]{James Worrell\texorpdfstring{\fnref{worrellthanks}}{}}



\address[tud]{Technische Universit\"at Dresden, Germany}
\address[oxf]{University of Oxford, United Kingdom}
\fntext[tudthanks]{The authors are supported by the DFG through
        the Collaborative Research Center 
        TRR 248 (see \url{https://perspicuous-computing.science}, project ID 389792660),
        the DFG project BA-1679/12-1,
        the Cluster of Excellence EXC 2050/1 (CeTI, project ID 390696704, as part of Germany's Excellence Strategy), and
        the Research Training Group QuantLA (GRK 1763). }
\fntext[kieferthanks]{Kiefer is supported by a University Research Fellowship of the Royal Society.}
\fntext[worrellthanks]{Worrell is supported by EPSRC grant EP/M012298/1.}
\maketitle              


\section{Introduction}

\begin{figure}[tbp]
\begin{minipage}{0.64\textwidth}%
\begin{tikzpicture}[initial text=, initial where=above, node distance= 0.18 \textwidth, scale=0.95, every node/.style={semithick, scale=0.95}, every path/.style={semithick, scale=0.95}]
    \node[state, ellipse, initial, minimum size=5 ex, inner sep=0.5 ex]  (q0)                        {\(q_0\)};
    \node[state, ellipse, minimum size=5 ex, inner sep=0.5 ex]           (qa)    [right of=q0]       {\(q_a\)};
    \node[state, ellipse, minimum size=5 ex, inner sep=0.5 ex, align=center]           (q1)    [right of=qa]       {\(q_1\)};
    \node[state, ellipse, minimum size=5 ex, inner sep=0 ex, align=center]           (q2)    [right of=q1]       {\(q_{k-2}\)};
    \node[align=center,draw=none]   (cdots) at($(q2)!0.5!(q1)$)   {\(\cdots\)};
    \node[state, ellipse, minimum size=5 ex, inner sep=0 ex]           (q3)    [right of=q2]       {\(q_{k-1}\)};
    \node[state, ellipse, accepting, minimum size=5 ex, inner sep=0.5 ex]           (qf)    [right of=q3]       {\(q_f\)};

    \draw[->]   (q0)    edge[loop below]     node[below]    {\(a,c\)}  (q0);
    \draw[->]   (q0)    edge[]               node[above]    {\(a\)}  (qa);
    \draw[->]   (qa)    edge[]               node[above]    {\(a,c\)}  (q1);
    \draw[->]   (q2)    edge[]               node[above]    {\(a,c\)}  (q3);
    \draw[->]   (q3)    edge[]               node[above]    {\(b\)}  (qf);
    \draw[->]   (qf)    edge[loop below]     node[below]    {\(a,b,c\)}  (qf);

    \draw[decorate,decoration={brace, mirror, amplitude=2 ex, raise=0.8 ex}, semithick] (qa.south) -- node[below, yshift=-3 ex] {\(k-1\) steps} (q3.south);
\end{tikzpicture}%
\end{minipage}%
\begin{minipage}{0.35\textwidth}%
\raggedleft
\hspace*{1 ex}
\begin{tikzpicture}[initial text=, initial where=above, node distance= 0.5 \textwidth, scale=0.95, every node/.style={semithick, scale=0.95}, every path/.style={semithick, scale=0.95}]
    \node[state, ellipse, initial, minimum size=5 ex, inner sep=0.5 ex, accepting]  (qa)                 {\(q_a\)};
    \node[state, ellipse, initial, minimum size=5 ex, inner sep=0.5 ex, accepting]  (qb) [right of=qa]   {\(q_b\)};

    \draw[->]   (qa)    edge[loop left]     node[left]    {\(a\)}  (qa);
    \draw[->]   (qa)    edge[bend left]     node[above]   {\(a\)}  (qb);
    \draw[->]   (qb)    edge[loop right]    node[right]   {\(b\)}  (qa);
    \draw[->]   (qb)    edge[bend left]     node[below]   {\(b\)}  (qa);
\end{tikzpicture}
\end{minipage}%
\caption{Left: UBA for ``eventually $b$ and $a$ appears $k$ steps before first
$b$'', right: A universal and separated UBA.}
\label{fig:uba-examples}
\end{figure}
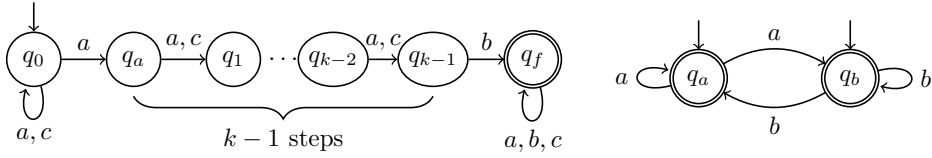

Unambiguity is a generalization of determinism that has been widely
studied in the theory and applications of
automata~\cite{Colcombet12,Colcombet15}.  In this paper we are
concerned with unambiguous automata over infinite words, that is,
nondeterministic automata in which every word has at most one
accepting run.  Our main results hold for the most commonly occurring
acceptance conditions (B\"{u}chi, Rabin, Muller, etc.), however in our
examples and experimental results we focus on the case of unambiguous
B\"{u}chi automata (UBA).  An example of a UBA is the automaton on the
right-hand side of Fig.~\ref{fig:uba-examples} in which both states
are initial and accepting.  This automaton is unambiguous by virtue of
the fact that there is exactly one run over every word.

Over infinite words, not only are UBA as expressive as
nondeterministic B\"uchi automata~\cite{Arnold85}, they can also be
exponentially more succinct than deterministic automata.  For example,
for a fixed $k\in \mathbb{N}$
the language ``eventually $b$ occurs and $a$ appears $k$ steps before
the first $b$'' over the alphabet $\{a,b,c\}$ is recognized by a UBA
with $k{+}1$ states (shown on the left-hand side of
Figure~\ref{fig:uba-examples}).  On the other hand, a deterministic
automaton for this language requires at least $2^k$ states, regardless
of the acceptance condition, as it needs to store the positions of the
$a$'s among the last $k$ input symbols.  Languages of this type arise
in a number of contexts, e.g., absence of unsolicited response in a
communication protocol---if a message is received, then it has been
sent in the recent past.

The exponential succinctness of UBA relative to deterministic automata
is also manifested in translations of linear temporal logic
(LTL) to automata.  The nondeterministic B\"{u}chi automata that are
obtained from LTL formulas by applying the classical closure algorithm
of~\cite{WVS83,VardiWolper86} are unambiguous. The generated automata
moreover enjoy the separation property: different states
have disjoint languages.  Thus, while the generation of deterministic
$\omega$-automata from LTL formulas incurs a double-exponential
blow-up in the worst case, the translation of LTL formulas into
separated UBA incurs only a single exponential blow-up. This fact has
been observed by several authors, see
e.g.~\cite{CouSahSut03,MorgensternThesis}, and adapted for LTL with
step parameters~\cite{Zimmermann13,ChaKat14}.
Besides allowing exactly one accepting run for every accepted word, there exist
weaker notions of unambiguity, where one allows a finite number of runs, or a
polynomial number of runs~\cite{RI89,HSS17}. These forms of unambiguity led to applications in
complementation of B\"uchi automata with a finite number of accepting runs~\cite{Rabinovich18}.


In the context of probabilistic model checking, UBA provide an elegant
alternative to deterministic automata for computing probabilities of
$\omega$-regular properties on finite-state Markov chains.  A
polynomial-time model checking procedure for UBA that represent safety
properties was given~\cite{BenLenWor14}, while~\cite{CouSahSut03}
gives a polynomial-time algorithm for separated UBA.  However,
separation is a strong restriction, as non-separated UBA (and even
DBA) can be exponentially more succinct than separated UBA,
see~\cite{BousLoed10}.  Furthermore, algorithms for the generation of
(possibly non-separated) UBA from LTL formulas that are more compact
than the separated UBA generated by the classical closure algorithm
have been realized in the tool \tulip{}
\cite{LenhardtThesis13,Len-Tulip13} and the automata library
\spot{}~\cite{Duret13}.  This motivates the design of algorithms that
operate with general UBA rather than the subclass of separated UBA.
For the analysis of finite-state Markov decision processes under
$\omega$-regular properties, there exist restricted forms of
nondeterministic automata such as limit-deterministic automata
\cite{Vardi85,CY95,SEJK16} or good-for-games automata
\cite{HP06,KMBK14}.

The main theoretical contribution of this paper is a polynomial-time
algorithm to compute the probability that the trajectory generated by
a given finite Markov chain satisfies an $\omega$-regular property
specified by a (not necessarily separated) unambiguous automaton.  We
present this algorithm under a mild assumption on the acceptance
condition of the automaton, which is satisfied by commonly occurring
acceptance conditions such as B\"{u}chi, Rabin, Muller, etc.
Furthermore we use our procedure to show that the model checking
problem of finite Markov chains against unambiguous automata lies in
the complexity class NC: the subclass of P comprising those problems
solvable in poly-logarithmic time by a parallel random-access machine
using polynomially many processors.

The existence of a polynomial-time algorithm for model checking Markov
chains against UBA has previously been claimed
in~\cite{BenLenWor13,BenLenWor14,LenhardtThesis13} (see
also~\cite{Len-Tulip13}).  However, these previous works share a
common fundamental error.  Specifically they rely on the claim that if
the language of a given UBA $\cA$ has positive probability with
respect to a Markov chain $\cM$, then there exists a state $s$ of
$\cM$ and a state $q$ of $\cA$ such that $q$ accepts almost all
trajectories emanating from $s$ (see~\cite[Lemma
7.1]{BenLenWor13},~\cite[Theorem 2]{BenLenWor14}%
\footnote{As the flaw is in the handling of the infinite behavior, the
  claim and proof of Lemma~1 in~\cite{BenLenWor14}, dealing with
  unambiguous automata over finite words, remain unaffected.  }
and~\cite[Section 3.3.1]{LenhardtThesis13}).  While this claim is true
in case $\cA$ is deterministic~\cite{CY95}, it need not hold when
$\cA$ is merely unambiguous (see
Remark~\ref{rem:counter-example-uba}).  We refer the reader
to~\cite{BKKKMW16} for full details and counterexamples to incorrect
claims in these works.

As a corollary of the above-mentioned NC bound we obtain another proof
of the fact that model checking LTL formulas on Markov chains can be
done in polynomial space (see~\cite{BusRubVar04} for a proof of this
fact using weak alternating automata and see~\cite{CY88,CY95} for an
automata-free approach).  Another corollary of our main result is that
one can decide in NC whether a given UBA accepts almost all words with
respect to the uniform distribution on $\omega$-words.  Recall that
while checking universality is known to be in NC for
deterministic B\"uchi automata and PSPACE-complete for NBA,
determining the complexity of the universality problem for UBA is a
long-standing open problem.  Polynomial-time procedures are only known
for separated UBA and other subclasses of
UBA~\cite{BousLoed10,IsaLoed12}.


A second contribution of our paper is an implementation of the new
algorithm as an extension of the model checker \prism{}, using the
automata library \spot{}~\cite{Duret13} for the generation of UBA from
LTL formulas and the \colt{} library~\cite{Hoschek04} for various
linear algebra algorithms.
We focus on unambiguous automata with a B\"uchi acceptance condition.
We evaluate our approach using the bounded
retransmission protocol case study from the PRISM benchmark suite
\cite{prismBenchmark} as well as specific aspects of our algorithm
using particularly ``challenging'' UBA.

We remark that there cannot exist a polynomial-time algorithm for model checking Markov decision processes (MDPs) against UBA.
This is because model checking LTL formulas on MDPs is 2EXPTIME-complete~\cite{CY95} and there is a single-exponential procedure for translating LTL formulas into UBA (cf.\ the proof of Corollary~\ref{cor:LTL-PSPACE}).

The rest of this article is structured as follows.
In Section~\ref{sec:prelim} we provide the necessary definitions and quote standard results on the spectral theory of nonnegative matrices.
In Section~\ref{sec:overview} we give an overview of our methodology.
Section~\ref{sec:uba} contains the main technical development with a polynomial-time model checking procedure.
Key technical lemmas are proved in Sections \ref{sec:recurrent-proof} and~\ref{sec:cut-proof}.
Section~\ref{sec:NC} improves the main result to an NC model checking procedure.
In Section~\ref{uba_implementation} we describe our implementation and experiments.
We conclude in Section~\ref{sec:conclusion}.

This article is a revised version of the CAV'16 conference
paper~\cite{BKKKMW16} and its extended version on
arxiv~\cite{cav16full}.  The main differences are that we present here
a direct proof technique, while \cite{BKKKMW16,cav16full} first
explain how to compute the measure of the language induced by strongly
connected UBA and then how to extend these techniques to arbitrary UBA
and the probabilistic model checking problem as discussed here.  In
contrast to \cite{BKKKMW16,cav16full} we consider here acceptance
conditions beyond B\"uchi acceptance.  Furthermore, the material on
experiments has been extended.

\section{Preliminaries}
\label{sec:prelim}

We assume the reader to be familiar with basic notions of Markov
chains and finite automata over infinite words, see, e.g., \cite{GraedelThomasWilke02,Kulkarni} and complexity theory, see, e.g., \cite{Pap94}.  In what follows, we
provide a brief summary of our notation for words, finite automata,
vectors and matrices, and Markov chains.
We also briefly summarize basic facts on the complexity class~NC and collect facts in the spectral theory of nonnegative matrices.

\paragraph*{Words}
Throughout the article, we suppose that $\Sigma$ is a finite non-empty alphabet.
For $L_1 \subseteq \Sigma^*$ and $L_2 \subseteq \Sigma^\omega$ we write $L_1 \cdot L_2$ for the concatenation of $L_1$ and~$L_2$, i.e., $L_1 \cdot L_2 = \{v w \in \Sigma^\omega: v \in L_1, \ w \in L_2\}$.
If $L_1 = \{v\}$ for some $v \in \Sigma^*$ then we may write $v L_2$ for $L_1 \cdot L_2$.

\paragraph*{Finite automata}
A (nondeterministic finite) automaton (over infinite words)
is a tuple $\cA = (Q,\Sigma,\delta,Q_0,\Acc)$
where $Q$ is the finite set of states, $Q_0 \subseteq Q$ is the set of initial
states, $\Sigma$ is the alphabet,
$\delta: Q \times \Sigma \to 2^Q$ is
the transition function, and $\Acc \subseteq 2^Q$ is the (Muller) acceptance condition.
We extend the transition function to $\delta: Q \times \Sigma^* \to 2^Q$
and to $\delta: 2^Q \times \Sigma^* \to 2^Q$ in the standard way.
Given states $q,r\in Q$ and a finite word
$w = a_0 a_1 \cdots a_{n-1} \in \Sigma^*$,
a \emph{run} for $w$ from $q$ to $r$ is a sequence
$q_0 q_1 \cdots q_n \in Q^{n+1}$ with $q_0=q$, $q_n=r$ and
$q_{i+1} \in \delta(q_i,a_i)$ for $i \in \{0, \ldots, n-1\}$.
A \emph{run} in $\cA$ for an infinite word
$w = a_0 a_2 a_3 \cdots \in \Sigma^{\omega}$ is an infinite sequence
$\rho = q_0 q_1 \cdots \in Q^{\omega}$ such that $q_0 \in Q_0$ and
$q_{i+1} \in \delta(q_i,a_i)$ for all $i \in \Nat$.  Run $\rho$ is
called \emph{accepting} if $\inf(\rho) \in \Acc$ where
$\inf(\rho) \subseteq Q$ is the set of states that occur infinitely
often in~$\rho$.  The \emph{language} $\cL(\cA)$ of accepted words
consists of all infinite words $w\in \Sigma^{\omega}$ that have at
least one accepting run.
If $R \subseteq Q$ then $\cA[R]$ denotes the automaton $\cA$ with $R$
as set of initial states.  If $\cA$ is understood from the context and
$q \in Q$ then we may write $\cL_q$ for~$\cL(\cA[\{q\}])$.
$\cA$ is called  \emph{deterministic} if $Q_0$ is a singleton and  $|\delta(q,a)| \leqslant 1$  for all $q \in Q$ and $a \in \Sigma$, and \emph{unambiguous} if each word $w\in \Sigma^{\omega}$ has at most one accepting run in~$\cA$.
Clearly, each deterministic automaton is unambiguous.

We assume that, given any set $R \subseteq Q$, one can compute in
polynomial time whether $R \in \Acc$.  This is the case, e.g., if
$\Acc$ is given as a \emph{B\"uchi} condition, i.e., as a set
$F \subseteq Q$ of \emph{accepting} states such that $R \in \Acc$ if
and only if $R \cap F \ne \emptyset$.
We use the acronym UBA for unambiguous B\"uchi automata.

We say that an automaton~$\cA$ has a \emph{diamond} from $q$ to~$r$
(where $q,r \in Q$) if there is a finite word $w \in \Sigma^*$ such
that there are at least two different runs for~$w$ from $q$ to~$r$.
If $\cA$ has no diamonds, we say that $\cA$ is \emph{diamond-free}.
If $\cA$ is an unambiguous automaton then one can make it diamond-free
in polynomial time and even in~NC: First remove all states that are unreachable
from~$Q_0$, along with their incoming and outgoing transitions.  Then
compute, in polynomial time, all states $q, r$ such that there is a
diamond from $q$ to~$r$.  By unambiguousness, we have
$\cL_r = \emptyset$, so we can remove~$r$ from~$\cA$, along with all
incoming and outgoing transitions, without changing $\cL(\cA)$.
Therefore we can and will generally assume that unambiguous automata are diamond-free.

\paragraph*{Complexity Theory}

Let $\{ C_n \}_{n\in\mathbb{N}}$ be a family of Boolean circuits such
that $C_n$ has $n$ input gates and any number of output gates.  Such a
family is said to be \emph{uniform} if there is a $\log n$-space
bounded Turing machine which on input $1^n$ outputs $C_n$.  Such a
family is moreover said to compute a function
$f:\{0,1\}^*\rightarrow\{0,1\}^*$ if for each $n\in \mathbb{N}$ and
every word $x \in \Sigma^n$ the output of $C_n$ on input $x$ equals
$f(x)$.
Function~$f$ is said to be computable in NC if there exists a
positive integer $d$ such that $f$ is computed by a uniform family
of circuits $\{ C_n \}_{n\in\mathbb{N}}$ where $C_n$ has depth
$O(\log^d n)$.  NC is widely considered as the subclass of
polynomial-time computable functions comprising those functions that
can be computed efficiently (i.e., in poly-logarithmic time) in
parallel (see, e.g.,~\cite[Chapter 15]{Pap94}).

A standard result of complexity theory is that a function computable
in NC is computable by a Turing machine using poly-logarithmic
space~\cite[Theorem 4]{Borodin77}.  Two facts that will be used below
are that reachability in directed graphs and matrix determinants (and
hence solving systems of linear equations) are both computable in
NC~\cite{Pap94}.

\paragraph*{Vectors and matrices}
We consider vectors and square matrices indexed by a finite set~$S$.
We use boldface for (column) vectors such as $\vec{v} \in \Real^S$,
and write $\vec{v}^\top$ for the transpose (a row vector)
of~$\vec{v}$.  The zero vector and the all-ones vector are denoted
by~$\vec{0}$ and $\vec{1}$, respectively.  A matrix
$M \in [0,1]^{S \times S}$ is called \emph{stochastic} if
$M \vec{1} = \vec{1}$, i.e., if every row of $M$ sums to one.  For a
set $U \subseteq S$ we write $\vec{v}_U \in \Real^U$ for the
restriction of~$\vec{v}$ to~$U$.  Similarly, for $T, U \subseteq S$ we
write $M_{T,U}$ for the submatrix of~$M$ obtained by deleting the rows
not indexed by~$T$ and the columns not indexed by~$U$.  The (directed)
\emph{graph} of a nonnegative matrix $M \in \Real^{S \times S}$ has
vertices $s \in S$ and edges $(s,t)$ if $M_{s,t} > 0$.  We may
implicitly associate~$M$ with its graph and speak about
graph-theoretic concepts such as reachability and strongly connected
components (SCCs) in~$M$.  For $s,t \in S$ we write
$\Paths_{s}(M) := \{s_0\cdots s_n \in sS^* : \bigwedge_{i=0}^{n-1}
M_{s_i,s_{i+1}}>0 \}$,
$\Paths_{s,t}(M) := \{s_0\cdots s_n \in \Paths_s(M) : s_n =t \}$, and
$\Paths^\omega_s(M) := \{s_0s_1 \cdots \in s S^\omega : \bigwedge_{i=0}^\infty M_{s_i,s_{i+1}}>0 \}$.


\paragraph*{Markov chains}
A (finite-state discrete-time) \emph{Markov chain} is a pair
$\cM = (S,M)$ where $S$ is the finite set of states, and
$M \in [0,1]^{S \times S}$ is a stochastic matrix that specifies
transition probabilities.  An \emph{initial distribution} is a
function $\iota : S \to [0,1]$ satisfying
$\sum_{s \in S} \iota(s) = 1$.  Such a distribution induces a
probability measure~$\Pr^{\cM}_\iota$ on the measurable subsets
of~$S^\omega$ in the standard way.
We may write $\Pr_\iota$ for~$\Pr^{\cM}_\iota$ if $\cM$ is understood.
If $\iota$ is concentrated on a
single state~$s$ then we may write $\Pr_s$
for~$\Pr_\iota$.  Note that
$\Pr_s(\Paths^\omega_s(M)) = 1$.

\paragraph*{Spectral Theory}
The \emph{spectral radius} of a matrix $M \in \Real^{S \times S}$,
denoted $\rho(M)$, is the largest absolute value of the eigenvalues
of~$M$.  The following result summarizes some facts in the spectral
theory of nonnegative matrices that will be used in the sequel.
In the
formulation below, we restrict attention to right eigenvectors.

\begin{theorem}
  Let $M\in \mathbb{R}^{S\times S}$ be a nonnegative matrix.  Then
  the following all hold:
\begin{enumerate}
\item The spectral radius $\rho(M)$ is an eigenvalue of~$M$ and there
  is a nonnegative eigenvector $\vec{x}$ with
  $M\vec{x}=\rho(M)\vec{x}$.
  Such a vector~$\vec{x}$ is called \emph{dominant}.
\item 
If $T \subseteq S$ then $\rho(M_{T,T}) \le \rho(M)$.
\item There is $C \subseteq S$ such that $M_{C,C}$ is strongly connected and $\rho(M_{C,C}) = \rho(M)$.
\end{enumerate}
\label{thm:nonnegative}
\end{theorem}

\begin{theorem}
  Let $M\in \mathbb{R}^{S\times S}$ be a strongly connected nonnegative
  matrix.  We have the following facts:
\begin{enumerate}
\item There is an eigenvector $\vec{x}$ with
  $M\vec{x}=\rho(M)\vec{x}$ such that $\vec{x}$ is
  strictly positive in all components.
\item The eigenspace associated with $\rho(M)$ is one-dimensional.
\item 
If $T \subsetneq S$ then $\rho(M_{T,T}) < \rho(M)$.
\item If $\vec{x} \geqslant 0$ and  $M\vec{x} \leqslant \rho(M)\vec{x}$
then $M\vec{x} = \rho(M)\vec{x}$.
\item If $M$ is strictly positive, i.e., $M_{i,j}>0$ for all $i,j \in S$, then $\lim_{i \to \infty} \left( M / \rho(M) \right)^i$ exists and is strictly positive.
\end{enumerate}
\label{thm:irreducible}
\end{theorem}

These results can mostly be found in~\cite[Chapter 2]{book:BermanP94}.
Specifically, Theorems \ref{thm:nonnegative}(1) and \ref{thm:irreducible}(1--2) are part of the Perron-Frobenius theorem, see \cite[Theorems 2.1.1, 2.1.4]{book:BermanP94};
Theorem~\ref{thm:nonnegative}(2) is \cite[Corollary 2.1.6(a)]{book:BermanP94};
Theorem~\ref{thm:nonnegative}(3) follows from \cite[Corollary 2.1.6(b)]{book:BermanP94};
Theorem~\ref{thm:irreducible}(3) follows from \cite[Corollary 2.1.6]{book:BermanP94};
Theorem~\ref{thm:irreducible}(4) follows from \cite[Corollary 2.1.11]{book:BermanP94};
Theorem~\ref{thm:irreducible}(5) follows from \cite[Theorem 8.2.7]{HornJohnson13}.

\section{Overview of the Methodology} \label{sec:overview}

Given a finite Markov chain $\cM$ and an unambiguous automaton $\cA$,
our goal is to compute the probability that a trajectory
generated by $\cM$ is accepted by~$\cA$ for a given initial distribution~$\iota$.  Suppose that $\cM$ has set
of states $S$ and $\cA$ has set of states $Q$.  Let vector
$\vec{z} \in \mathbb{R}^{Q\times S}$ be such that for each $q\in Q$
and $s\in S$, $\vec{z}_{\<qs\>}$ is the probability that a trajectory
of $\cM$ starting in state $s$ is accepted by $\cA$ starting in state
$q$.  It suffices to compute $\vec{z}$.

Our strategy to compute $\vec{z}$ is to find a system of linear
equations that has $\vec{z}$ as unique solution.  To this end, the
first step is to form the product of the transition functions of $\cA$
and $\cM$, thereby obtaining a nonnegative matrix
$B \in \mathbb{R}^{(Q\times S)\times (Q\times S)}$, and then to show that
  $B\vec{z}=\vec{z}$.  However this system of linear equations, being
  homogeneous, certainly does not determine $\vec{z}$ uniquely.

In case $\cA$ is deterministic, it is relatively straightforward to
write down extra linear equations that pin down $\vec{z}$ uniquely:
one looks at the directed graph underlying matrix $B$ and classifies
the bottom SCCs of this graph as being either accepting or rejecting,
according to the automaton states that appear in them.  One then adds
an equation $\vec{z}_{\<q s\>}=1$ for every pair $(q,s)$ in an accepting
  bottom SCC and an equation $\vec{z}_{\<q s\>}=0$ for every pair $(q,s)$
    in a rejecting bottom SCC.

In order to generalize the above analysis to unambiguous automata we
rely extensively on the spectral theory of nonnegative matrices.  In
particular, rather than looking at bottom SCCs of matrix $B$ we focus
on SCCs that induce submatrices of $B$ with spectral radius one.  We
call the latter \emph{recurrent SCCs}.  Recurrent SCCs need not be
bottom SCCs when $\cA$ is unambiguous. We classify recurrent SCCs as
being accepting or rejecting in similar manner to the deterministic
case.  For each pair $(q,s)$ in a rejecting recurrent SCC we have an
equation $\vec{z}_{\<qs\>}=0$.  For an accepting recurrent SCC
$D\subseteq Q\times S$ we do not in general have $\vec{z}_{\<qs\>}=1$
for each $(q,s) \in D$.  Rather, writing $\vec{z}_D$ for the
restriction of $\vec{z}$ to $D$, we have $\vec{\mu}^\top \cdot \vec{z}_D=1$
for some weight vector $\vec{\mu} \in [0,1]^D$.  While such a weight
vector can be computed from the determinization of $\cA$, we show how
to compute a weight vector in polynomial time (and even NC) by
exploiting structural properties of unambiguous automata.  Given such
a weight vector $\vec{\mu}$ we add the single linear equation
$\vec{\mu}^\top \cdot \vec{z}_D=1$ to our system and thereby ensure a
unique	solution.

\subsection{Structure of Section~\ref{sec:uba}}
The following section, Section~\ref{sec:uba}, follows this approach to prove our main result, Theorem~\ref{thm:PMC-MC-UBA}.
Section~\ref{sub:linear-system} sets up the basic linear system, see Lemma~\ref{lem:basic-eq}.
The linear system can be written in matrix form as $\vec{\zeta} = B \vec{\zeta}$, see~\eqref{eq:basic-eq-2-matrix}, where $\vec{\zeta}$ is a vector of variables.
This system is satisfied by the vector~$\vec{z}$ which contains the probabilities in question; i.e., we have $\vec{z} = B \vec{z}$.
In Section~\ref{sub:linear-system}, specifically in Proposition~\ref{prop:powers-of-B}, we start deriving further properties of the matrix~$B$.
Matrix~$B$ can be viewed as a weighted graph representing a product of the automaton and the Markov chain.
This dual view of~$B$ (on the one hand containing coefficients of the linear system, on the other representing the product as a weighted graph) drives the technical development in the rest of Section~\ref{sec:uba}.

In Section~\ref{sub:recurrent} we first show, in Proposition~\ref{prop:spectral-radius-le-1}, that the spectral radius of~$B$ is at most~$1$.
Recurrent SCCs of~$B$ are defined to be those where the corresponding submatrix of~$B$ has spectral radius exactly~$1$.
Recurrent SCCs are the analogues of bottom SCCs in the product of a deterministic automaton and a Markov chain.
In Lemma~\ref{lem:recurrent} (proved in Section~\ref{sec:recurrent-proof}) we provide crucial properties of recurrent SCCs.
Specifically we show that a coordinate of~$\vec{z}$ is strictly positive if and only if it belongs to an accepting recurrent SCC (where accepting means that the set of automaton states associated with the SCC is accepting).

As mentioned above, the system $\vec{\zeta} = B \vec{\zeta}$ does not uniquely determine~$\vec{z}$.
Therefore, for each accepting recurrent SCC~$D$, we add an equation $\vec{\mu}_D^\top \vec{\zeta}_D = 1$, where $\vec{\mu}_D \in [0,1]^D$ is a weight vector which we call $D$-\emph{normalizer}.
We show in Section~\ref{sub:cut} that such normalizers can be taken as the characteristic vectors of certain subsets of~$D$ called \emph{cuts}.
Again we benefit from a dual view: on the one hand cuts provide the necessary normalizing equations, on the other hand they describe a combinatorial property: intuitively, a cut is a minimal subset of~$D$ that cannot be ``driven extinct'' by the Markov chain.
Lemma~\ref{lem:cut} shows important properties of cuts, including their polynomial-time computability.
The algorithm (given in Section~\ref{sec:cut-proof}) is purely combinatorial and heavily exploits the diamond-freeness of the automaton.

With the necessary ingredients at hand, in Section~\ref{sub:augmented} we give the full linear system, pinning down~$\vec{z}$ uniquely; see Lemma~\ref{lem:linear-system}.
Then we prove the main theorem, Theorem~\ref{thm:PMC-MC-UBA}, by giving the overall algorithm.

\section{A Polynomial-Time Model Checking Procedure}
\label{sec:uba}

Given a Markov chain~$\cM$, an initial distribution~$\iota$, and an
automaton~$\cA$ whose alphabet is the state space of~$\cM$, the
\emph{probabilistic model-checking problem} is to compute
$\Pr_\iota(\cL(\cA))$.  This problem is solvable in polynomial
time in case $\cA$ is a deterministic automaton and in polynomial space for nondeterministic automata~\cite{CY95,BusRubVar04}.
The main result of this article
extends the polynomial-time bound from deterministic automata to
unambiguous automata.

\begin{theorem}
  \label{thm:PMC-MC-UBA}
  Given a Markov chain $\cM$, an initial distribution~$\iota$, and an
  unambiguous automaton~$\cA$, the value $\Pr_\iota(\cL(\cA))$
  is computable in polynomial time.
\end{theorem}

\begin{remark}
\label{rem:counter-example-uba}
The statement of Theorem~\ref{thm:PMC-MC-UBA} has already been presented
in \cite{BLW13} (see also \cite{LenhardtThesis13,BenLenWor14}). However, the
presented algorithm is flawed.
The error stems from the incorrect claim that if $\Pr_\iota(\cL(\cA))$
is strictly positive then there is necessarily a state $q$ of $\cA$
and state $s$ on $\cM$ such that $q$ accepts almost all trajectories
emanating from $s$.
A counterexample is obtained by taking
$\cA$ to be the automaton on the right in Fig.~\ref{fig:uba-examples}
and $\cM$ the Markov chain that generates the uniform distribution on
$\{a,b\}^\omega$.  Clearly state $q_a$ of $\cA$ accepts all words that begin
with $a$, while state $q_b$ accepts all words that begin with $b$.
Thus $\cA$ is universal and its language has probability~$1$ under the
uniform distribution.  However each of the two languages $\cL_{q_a}$ and~$\cL_{q_b}$
has probability $\frac{1}{2}$ under the uniform distribution.
\end{remark}

The remainder of this section gives a proof of
Theorem~\ref{thm:PMC-MC-UBA}.  The development heavily relies on two
technical lemmas (Lemmas~\ref{lem:recurrent} and~\ref{lem:cut}), whose
proofs are given in Sections \ref{sec:recurrent-proof}
and~\ref{sec:cut-proof}, respectively.

\subsection{The Basic Linear System} \label{sub:linear-system}

Let $\cM = (S,M)$ be a Markov chain, $\iota$ an initial distribution,
and $\cA = (Q,S,\delta,Q_0,\Acc)$ an unambiguous automaton.
\begin{lemma} \label{lem:basic-eq}
The following equations hold:
\begin{align}
\Pr_\iota( \cL(\cA))
\ \  &=  \ \
\sum_{s\in S} \ \iota(s) \cdot \sum_{q\in Q_0} \Pr_s(\cL_q)  \label{eq:basic-eq-1} \\
\text{for all }q \in Q \text{ and } s \in S : \ \ \Pr_s(\cL_q)
\ \ &= \ \
\sum_{t\in S} \sum_{r \in \delta(q,s)} M_{s,t}\, \cdot\, \Pr_t(\cL_r) \label{eq:basic-eq-2}
\end{align}
\end{lemma}
\begin{proof}
For all $q \in Q$ and $s \in S$ we have
$
\cL_q \cap s S^\omega
= s \bigcup_{r \in \delta(q,s)} \cL_r
$.
Hence:
\begin{align*}
\Pr_s(\cL_q)
&\ = \ \Pr_s(\cL_q \cap s S^\omega)
\ = \ \Pr_s\Big(s \bigcup_{r \in \delta(q,s)} \cL_r\Big) \\
&\ = \ \sum_{t \in S} M_{s,t} \cdot \Pr_t\Big(\bigcup_{r \in \delta(q,s)}\cL_r\Big)
\end{align*}
Since $\cA$ is unambiguous, the sets $\cL_r$ are pairwise disjoint.
Hence \eqref{eq:basic-eq-2} follows.
Equation~\eqref{eq:basic-eq-1} is shown similarly.
\qed
\end{proof}

Define the vector $\vec{z} \in [0,1]^{Q \times S}$ by
$\vec{z}_{\<q s\>} = \Pr_s(\cL_q)$.  By
Lemma~\ref{lem:basic-eq}, to prove
Theorem~\ref{thm:PMC-MC-UBA} it suffices to compute $\vec{z}$ in
polynomial time.  To this end, define a square matrix
$B \in [0,1]^{(Q \times S) \times (Q \times S)}$ by
\begin{gather}
 B_{\<q s\>,\<r t\>}
   \ \ = \ \
  \begin{cases}
     M_{s,t} & \text{if } r\in \delta(q,s) \\[0.5ex]
     0 & \text{otherwise.}
  \end{cases}
\label{eq:defB}
\end{gather}
As $\cA$ is unambiguous, $B$ need not to be a stochastic matrix, whereas if
$\cA$ is deterministic, $B$ is a stochastic matrix.
By~\eqref{eq:basic-eq-2} we have $\vec{z} = B \vec{z}$, i.e., $\vec{z}$~solves the system of linear equations
\begin{equation}
 \vec{\zeta} = B \vec{\zeta}\,, \label{eq:basic-eq-2-matrix}
\end{equation}
where $\vec{\zeta}$ is a vector of variables indexed by $Q \times S$.

\begin{figure}
\begin{center}
\begin{tabular}{cc}
\begin{tikzpicture}[->,>=stealth',shorten >=1pt,auto,node distance=2.5cm,
                    semithick]
  \tikzstyle{every state}=[draw=black,text=black]

  \node[initial,state] (A)                    {$q_0$};
  \node[state]         (B)       [right of=A] {$q_1$};
  \node[accepting,state]         (D)       [below of=B] {$q_2$};

  \path (A) edge [bend left] node {$a$} (B);
  \path (B) edge [bend left] node[swap] {$a$} (A);
  \path (B) edge [bend left] node[swap] {$a$} (D);
  \path (D) edge node {$b$} (A);
  \path (D) edge [bend left] node {$b$} (B);
  \path (D) edge [loop below] node {$b$} (D);
\end{tikzpicture}
\qquad\mbox{}
&
\begin{tikzpicture}[->,>=stealth',shorten >=1pt,auto,node distance=2.5cm,
                    semithick]
  \tikzstyle{every state}=[draw=black,text=black]

\node[state] (A) {$a$};
\node[state] (B) [right of=A] {$b$};

\path (A) edge [bend left] node {$\frac{1}{2}$} (B);
\path (B) edge [bend left] node {$\frac{1}{2}$} (A);
\path (A) edge [loop above] node {$\frac{1}{2}$} (A);
\path (B) edge [loop above] node {$\frac{1}{2}$} (B);
\end{tikzpicture}
\\
UBA~$\cA$ & Markov chain~$\cM$
\end{tabular}
\\[20mm]
\begin{tabular}{c}
\begin{tikzpicture}[->,>=stealth',shorten >=1pt,auto,node distance=2.5cm,
                    semithick]
  \tikzstyle{every state}=[draw=black,text=black]

  \node[state,inner sep=2pt,scale=0.98]         (A)                    {$\<q_0a\>$};
  \node[state,inner sep=2pt,scale=0.98]         (B)       [right of=A] {$\<q_1a\>$};
  \node[state,inner sep=2pt,scale=0.98]         (D)       [below of=B] {$\<q_2b\>$};
  \node[state,inner sep=2pt,scale=0.98]         (E)       [below of=A] {$\<q_1b\>$};
    \node[state,inner sep=2pt,scale=0.98]       (F)       [right of=B] {$\<q_0b\>$};
    \node[state,inner sep=2pt,scale=0.98]       (H)       [right of=D] {$\<q_2a\>$};

  \path (A) edge [bend left] node {$\frac{1}{2}$} (B);
  \path (B) edge [bend left] node[swap] {$\frac{1}{2}$} (A);
  \path (B) edge [bend left] node[swap] {$\frac{1}{2}$} (D);
  \path (D) edge node[xshift=2mm] {$\frac{1}{2}$} (A);
  \path (D) edge [bend left] node {$\frac{1}{2}$} (B);
   \path (D) edge [loop below] node {$\frac{1}{2}$} (D);
  \path (A) edge node {$\frac{1}{2}$} (E);
  \path (B) edge node[swap] {$\frac{1}{2}$} (F);
  \path (D) edge node[swap] {$\frac{1}{2}$} (H);

\path  (D) edge node {$\frac{1}{2}$} (E);
\path  (D) edge node[swap,xshift=-2mm] {$\frac{1}{2}$} (F);
\draw [->,rounded corners] (B) -- ++(0,0.7) -- node {$\frac12$} ++(3.3,0) -- ++(0,-3.2) -- (H);
\end{tikzpicture}
\\
product~$B$
\end{tabular}
\end{center}
\caption{A UBA~$\cA$, Markov chain $\cM$, and their product
  $B$, as described in Example~\ref{ex:product}.}
\label{fig:product}
\end{figure}
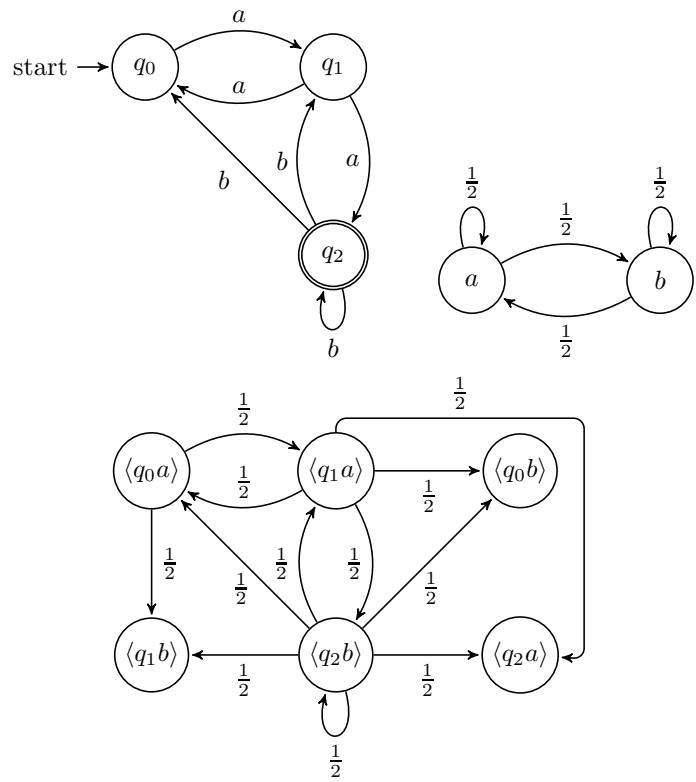

\begin{example}
Consider the UBA~$\cA$, shown above left in
Figure~\ref{fig:product}.  This automaton is unambiguous since any two
distinct states have disjoint languages:
\begin{align*}
\cL_{q_0} & \ = \ \{ a^{2 m} w : w \in (b^+a^*)^\omega,\, m \geqslant 1 \}\\
\cL_{q_1} & \ = \ \{ a^{2 m - 1} w : w \in (b^+a^*)^\omega,\, m \geqslant 1 \} \\
\cL_{q_2} & \ = \ \{ w : w \in (b^+a^*)^\omega \}
\end{align*}
Consider moreover the
two-state Markov chain $\cM$, shown above right in
Figure~\ref{fig:product}.  The weighted graph on the bottom of
Figure~\ref{fig:product} represents the matrix $B$, obtained from
$\cA$ and $\cM$ according to Equation \eqref{eq:defB}.  It is natural
to think of $B$ as a product of $\cA$ and $\cM$.  Notice that $B$ is
not stochastic: the sum of the entries in each row (equivalently, the
total outgoing transition weight of a graph node) may be strictly less
than one and may be strictly greater than one.
\label{ex:product}
\end{example}

Although \eqref{eq:basic-eq-2-matrix} contains one equation for each
$\<q s\> \in Q \times S$, the vector~$\vec{z}$ is not necessarily the
unique solution of~\eqref{eq:basic-eq-2-matrix}, e.g., any scalar
multiple of $\vec{z}$ is also a solution.  Below we identify suitable
equations that, together with~\eqref{eq:basic-eq-2-matrix}, have
$\vec{z}$ as a unique solution.  These extra equations are based on
analysis of the strongly connected components (SCCs) of matrix $B$.

The following proposition shows that the powers of (submatrices of) $B$ admit a probabilistic interpretation.
Roughly, Proposition~\ref{prop:powers-of-B} states that the entries of the $n$th matrix power are probabilities of length-$n$ paths in the Markov chain that trigger certain runs in the automaton.
\begin{proposition} \label{prop:powers-of-B}
Let $C \subseteq Q \times S$ and $\<q s\>, \<r t\> \in C$.
Let $n \in \Nat$.
Define $A := B_{C,C}$ and
\[
E^{C,n}_{\<q s\>, \<r t\>} \ := \ \{s_0 s_1 \cdots \in s S^\omega : \exists\, q_1 \cdots q_n\,.\, \<q_0 s_0\> \cdots \<q_n s_n\> \in \Paths_{\<q s\>, \<r t\>}(A)\}\,.
\]
Then $(A^n)_{\<q s\>, \<r t\>} = \Pr_s\big(E^{C,n}_{\<q s\>, \<r t\>}\big)$.
In particular:
\begin{equation*}
(B^n)_{\<q s\>, \<r t\>} \ = \ \Pr_s(\{s_0 s_1 \cdots \in s S^\omega: r \in \delta(q,s_0 \cdots s_{n-1}),\ s_n=t\})
\end{equation*}
\end{proposition}
\begin{proof}
Let $n \in \Nat$ and $s_0 \cdots s_n \in \Paths_{s,t}(M)$.
Since the unambiguous automaton~$\cA$ is diamond-free, it has at most one run for $s_0 \cdots s_{n-1}$ from $q$ to~$r$.
A sequence $q_0 \cdots q_n$ is such a run if and only if $\<q_0 s_0\> \cdots \<q_n s_n\> \in \Paths_{\<q s\>, \<r t\>}(B)$.
Such a path is in $\Paths_{\<q s\>, \<r t\>}(A)$ if and only if $\<q_0 s_0\>, \ldots, \<q_n s_n\> \in C$.
In that case we have:
\[
 \Pr_s(s_0 \cdots s_n S^\omega) \ = \ M_{s_0, s_1} \cdot \ldots \cdot M_{s_{n-1}, s_n} \ = \ A_{\<q_0 s_0\>, \<q_1 s_1\>} \cdot \ldots \cdot A_{\<q_{n-1} s_{n-1}\>, \<q_n s_n\>}
\]
We conclude that, for fixed $\<q s\>, \<r t\>$, the probability of those $s_0 \cdots s_n \in \Paths_{s,t}(M)$ for which there are $q_0 \cdots q_n$ with  $\<q_0 s_0\> \cdots \<q_n s_n\> \in \Paths_{\<q s\>, \<r t\>}(A)$ equals $(A^n)_{\<q s\>, \<r t\>}$.
The proposition follows.
\qed
\end{proof}

\subsection{Recurrent SCCs}
\label{sub:recurrent}
In this section we define a notion of recurrent SCC for the matrix
$B$, generalizing the familiar notion of the same name for finite
Markov chains.  We classify each recurrent SCC $D$ as either accepting
or non-accepting, showing that $\vec{z}_D$ is nonzero just in case
$D$ is accepting.

If automaton $\cA$ is deterministic then the matrix $B$ in
\eqref{eq:basic-eq-2-matrix} is stochastic.  While $B$ need not be
stochastic if $\cA$ is merely unambiguous, we have
\begin{proposition} \label{prop:spectral-radius-le-1}
  $\rho(B) \leqslant 1$.
\end{proposition}
\begin{proof}
By Proposition~\ref{prop:powers-of-B}, for all $n$, all entries of~$B^n$ are at most~$1$.
Let $\vec{x} \ne \vec{0}$ be a dominant eigenvector, i.e., $B \vec{x} = \rho(B) \vec{x}$.
Then $\rho(B)^n \vec{x} = B^n \vec{x}$ is bounded over all~$n \geqslant 0$.
It follows that $\rho(B) \leqslant 1$.
\qed
\end{proof}

Using Theorem~\ref{thm:nonnegative}(2), it follows from
Proposition~\ref{prop:spectral-radius-le-1} that for any
$D \subseteq Q \times S$ we have $\rho(B_{D,D}) \leqslant 1$.  An SCC
$D \subseteq Q \times S$ of $B$ is called \emph{recurrent} if
$\rho(B_{D,D}) = 1$.  Such an SCC~$D$ is said to be \emph{accepting} if
$\{ q : \exists\, s\,.\, \<q s\> \in D \} \in Acc$, i.e., if the collection
of all automaton states in $D$ is an accepting set.

The following lemma summarizes the key properties of recurrent SCCs
that we will need.
\begin{restatable}{lemma}{lemrecurrent}\label{lem:recurrent}\label{LEM:RECURRENT}
  Let $D$ be a recurrent SCC.
\begin{enumerate}
\item We have $\vec{z}_D = B_{D,D} \vec{z}_D$.
\item For all $d \in D$, we have $\vec{z}_d > 0$ iff $D$ is accepting.
\end{enumerate}
\end{restatable}
\noindent Lemma~\ref{lem:recurrent} is proved in Section~\ref{sec:recurrent-proof}.

\begin{example}
Consider the matrix $B$ from Example~\ref{ex:product}, shown in
Figure~\ref{fig:product}.  This matrix has a single recurrent SCC,
namely $D = \{\<q_0a\>,\<q_1a\>,\<q_2b\>\}$.
It is accepting, as the automaton uses a B\"uchi acceptance condition.
We have:
\[
B_{D,D} \ = \
\begingroup 
\setlength\arraycolsep{6pt}
\begin{pmatrix}
0 & 1/2 & 0 \\
1/2 & 0 & 1/2 \\
1/2 & 1/2 & 1/2
\end{pmatrix}
\endgroup
\]
The vector $(1 \ \ 2 \ \ 3 )^\top$ is a dominant eigenvector of $B_{D,D}$.
\end{example}

\subsection{Cuts}
\label{sub:cut}
In this subsection we introduce the notion of a cut of an SCC.
Among other things, this yields a purely graph-theoretic
characterization of recurrent SCCs: an SCC is recurrent just in case
it has a cut.

Let $D \subseteq Q \times S$ be an SCC of~$B$.  A set
$\alpha \subseteq D$ is called a \emph{fiber} of $D$ if it can be
written $\alpha=\alpha'\times\{s\}$ for some $\alpha'\subseteq Q$ and
$s\in S$.  Given such a fiber $\alpha$ and $t \in S$, if $M_{s,t}>0$
then we define a fiber
\[ \alpha\then t := \{ \<q t\> \in D : q \in \delta(\alpha',s) \} \,
  . \] If $M_{s,t}=0$ then $\alpha\then t$ is undefined.  We extend
this definition inductively by
$\alpha \then \varepsilon := \alpha$ and
$\alpha \then w t := (\alpha\then w)\then t$ for $t \in S$ and
$w \in S^*$.
We have that $\alpha \then w$ is defined if and only if $s w$ describes a path in~$M$.
If $\alpha$ is a singleton~$\{d\}$, we may write $d \then w$ for $\alpha \then w$.

  We call a fiber $\alpha \subseteq D$ a \emph{cut} of~$D$ if
  (i)~$\alpha = d \then v$ for some $d \in D$ and $v \in S^*$,
  and (ii)~$\alpha\then w \neq \emptyset$ holds for all $w\in S^*$ such that
  $\alpha\then w$ is defined.
Clearly if $\alpha$ is a cut then so is $\alpha\then w$ when the
latter is defined.  Given a cut $\alpha\subseteq D$, we call its
characteristic vector $\vec{\mu} \in \{0,1\}^D$ a \emph{cut vector}.

The following lemma summarizes the key properties of cuts that we will need.%
\begin{restatable}{lemma}{lemcut}\label{lem:cut}\label{LEM:CUT}
Let $D$ be an SCC.  Then
\begin{enumerate}
\item $D$ is recurrent if and only if it has a cut.
\item If $D$ is accepting recurrent and $\vec{\mu}$ is a cut vector then $\vec{\mu}^\top \vec{z}_D=1$.
\item If $D$ is recurrent one can compute a cut in polynomial time.
\end{enumerate}
\end{restatable}
\noindent Lemma~\ref{lem:cut} is proved in Section~\ref{sec:cut-proof}.

Given an accepting recurrent SCC~$D$, say that a vector $\vec{\mu} \in [0,1]^D$ is a
\emph{$D$-normalizer} if $\vec{\mu}^\top \vec{z}_D = 1$.  Then
Lemma~\ref{lem:cut}(2) says that a cut vector for $D$ is a
$D$-normalizer.

\begin{example}
  Consider the matrix $B$ from Example~\ref{ex:product}, shown in
  Figure~\ref{fig:product}, and its single recurrent SCC $D = \{\<q_0a\>,\<q_1a\>,\<q_2b\>\}$.  Then
  $\<q_0 a\> \then ab = \{ \<q_2b\> \}$ is a cut and
  $\<q_2 b \> \then a = \{ \<q_0a\>,\<q_1a\>\}$ is another cut.  Both
  the associated cut vectors are $D$-normalizers.
  For  example, let $\vec{\mu}$ be the cut vector associated with $\{ \<q_0a\>,\<q_1a\>\}$, i.e., $\vec{\mu}^\top = (1 \ \ 1 \ \ 0)$.
  Since $\vec{z}_D = (1/3 \ \ 2/3 \ \ 1)^\top$, we have $\vec{\mu}^\top \vec{z}_D = 1$.
\end{example}

\subsection{The Augmented Linear System} \label{sub:augmented}
We now extend \eqref{eq:basic-eq-2-matrix} to a linear
system that has $\vec{z}$ as unique solution.
\begin{lemma} \label{lem:linear-system}
Let $\cD_+$ be the set of accepting recurrent SCCs, and $\cD_0$ the
set of non-accepting recurrent SCCs.  For each $D \in \cD_+$ let
$\vec{\mu}_D \in [0,1]^D$ be a $D$-normalizer (which
exists by Lemma~\ref{lem:cut}(2)).  Then $\vec{z}$~is the unique
solution of the following linear system:
\begin{equation}
\begin{aligned}
&&\vec{\zeta} & = B \vec{\zeta} \\
\text{for all } D \in \cD_+ :&& \quad \vec{\mu}_D^\top \vec{\zeta}_D & = 1 \\
\text{for all } D \in \cD_0 :&& \quad \vec{\zeta}_D & = \vec{0}
\end{aligned}
\label{eq:linear-system}
\end{equation}
\end{lemma}
\begin{proof}
  The vector~$\vec{z}$ solves~\eqref{eq:linear-system}: indeed, this
  follows from the equality $\vec{z} = B \vec{z}$, the definition of a
  $D$-normalizer, and Lemma~\ref{lem:recurrent}(2).

  It remains to show uniqueness.  To this end, let
  $\vec{x}$ solve~\eqref{eq:linear-system}.  We show that
  $\vec{x} = \vec{z}$.
We proceed by induction over the DAG of SCCs of~$B$.
Let $D \subseteq Q \times S$ be any SCC (possibly trivial, i.e., $D = \{d\}$ for some $d \in Q \times S$, and $B_{d,d} = 0$).
Let us write $D{\downarrow}$ for the set of SCCs directly below~$D$.
By the induction hypothesis, we have $\vec{x}_C = \vec{z}_C$ for all SCCs $C \in D{\downarrow}$.
We have to show that $\vec{x}_D = \vec{z}_D$.
Since $\vec{x} = B \vec{x}$ and $\vec{z} = B \vec{z}$, we have
\begin{equation} \label{eq:linear-system-uniqueness}
\begin{aligned}
\vec{x}_D - \vec{z}_D \ &= \ B_{D, Q \times S} (\vec{x} - \vec{z}) \ = \ B_{D,D} (\vec{x}_D - \vec{z}_D) + \sum_{C \in D{\downarrow}} B_{D,C} (\vec{x}_C - \vec{z}_C) \\
                        &= \ B_{D,D} (\vec{x}_D - \vec{z}_D) \quad \text{by the induction hypothesis.}
\end{aligned}
\end{equation}

\begin{itemize}
\item
Let $D$ be non-recurrent.
Then we must have $\vec{x}_D = \vec{z}_D$, as otherwise, by~\eqref{eq:linear-system-uniqueness}, the vector $\vec{x}_D - \vec{z}_D$ would be an eigenvector of~$B_{D,D}$ associated with eigenvalue~$1$, implying $\rho(B_{D,D}) \ge 1$, and thus contradicting the assumption that $D$ is not recurrent.
\item
Let $D$ be recurrent. If $D \in \cD_0$, then $\vec{x}_D = \vec{0} = \vec{z}_D$.
Therefore, we can assume that $D \in \cD_+$.
By Lemma~\ref{lem:recurrent}(1), $\vec{z}_D = B_{D,D} \vec{z}_D$.
Thus, with~\eqref{eq:linear-system-uniqueness}, $\vec{x}_D = B_{D,D} \vec{x}_D$.
By Theorem~\ref{thm:irreducible}(2), the eigenspace of~$B_{D,D}$ associated with the spectral radius is one-dimensional, implying that $\vec{x}_D$ is a scalar multiple of~$\vec{z}_D$.
We have $\vec{\mu}_D^\top \vec{x}_D = 1 = \vec{\mu}_D^\top \vec{z}_D$, hence $\vec{x}_D = \vec{z}_D$.
\qed
\end{itemize}
\end{proof}

Now we can prove our main result, Theorem~\ref{thm:PMC-MC-UBA}.
\begin{proof}[of Theorem~\ref{thm:PMC-MC-UBA}]
Given a Markov chain~$\cM$, an initial distribution~$\iota$, and a diamond-free unambiguous automaton~$\cA$, proceed as follows.
\begin{enumerate}
\item Set up the matrix~$B$ from Section~\ref{sub:linear-system}.
\item Compute the SCCs of~$B$.
\item For any SCC $C$, check whether $C$ is recurrent, by seeing if the linear system $B_{C,C}\vec{x}=\vec{x}$ has a nonzero solution.
\item For any accepting recurrent SCC~$D$, compute a cut vector $\vec{\mu}$ using Lemma~\ref{lem:cut}(3).
\item Solve the linear system~\eqref{eq:linear-system} in Lemma~\ref{lem:linear-system}.
\item Compute $\Pr_\iota(\cL(\cA))$ using~\eqref{eq:basic-eq-1} in Lemma~\ref{lem:basic-eq}. \qed
\end{enumerate}
\end{proof}

\section{Proof of Lemma~\ref{lem:recurrent}}
\label{sec:recurrent-proof}

In this section we prove Lemma~\ref{lem:recurrent}, which is restated here.

\lemrecurrent*

\subsection{Proof of Lemma~\ref{lem:recurrent}(1)}
\paragraph{Proof of Lemma~\ref{lem:recurrent}(1)}
Recall that $\vec{z} = B \vec{z}$.
Thus, $\vec{z}_D = B_{D,Q\times S} \vec{z} \geqslant B_{D,D} \vec{z}_D$.
Since $\rho(B_{D,D}) = 1$ and $D$ is strongly connected, it follows, by Theorem~\ref{thm:irreducible}(4), that $\vec{z}_D = B_{D,D} \vec{z}_D$.  \qed

\subsection{Proof of Lemma~\ref{lem:recurrent}(2)}
The proof of Lemma~\ref{lem:recurrent}(2) requires some auxiliary
definitions and results.

Let $C, D \subseteq Q \times S$ be two SCCs of matrix $B$.  We write
$C \preceq D$ if $C$ is reachable from~$D$.  Note that
$\mathord{\preceq}$ is a partial order on the SCCs.
In case matrix $B$ is stochastic the recurrent SCCs are just the
bottom SCCs.  While this does not hold in general, we have the
following result.
\begin{proposition}
\label{prop:lower-matrices-zero}
Let $D \subseteq Q \times S$ be a recurrent SCC.  Then all SCCs~$C \prec D$
are such that $\vec{z}_C = \vec{0}$.
\end{proposition}
\begin{proof}
Recall that $B \vec{z} = \vec{z}$.
Thus, $B_{D,Q\times S} \vec{z} = \vec{z}_D = B_{D,D} \vec{z}_D$ by Lemma~\ref{lem:recurrent}(1).
Hence, we have for any $c \in (Q \times S) \setminus D$ that $B_{D,c} \vec{z}_c = \vec{0}$.
So if $B_{D,c} \ne \vec{0}$ then $\vec{z}_c = 0$.
It follows that $\vec{z}_C = \vec{0}$ for
all SCCs $C$ directly below~$D$.
Using the definitions of~$\vec{z}$ and~$B$ it follows that $\vec{z}_C = \vec{0}$ must hold for all SCCs~$C \prec D$.
\qed
\end{proof}

In the following two propositions we consider paths in~$M$ along with ``corresponding'' paths in~$B$.
The gist of Propositions \ref{prop:recurrent} and~\ref{prop:recurrent-2} is that paths in~$M$ that have ``corresponding'' paths in~$B$ that linger in a strict subset of an SCC are negligible.

Let $\pi_Q : Q\times S \rightarrow Q$ and $\pi_S : Q\times S
\rightarrow S$ denote the obvious projection maps.  We extend both
maps pointwise to finite and infinite sequences; e.g., we have $\pi_Q
: (Q\times S)^* \rightarrow Q^*$.

\begin{proposition}
  Let $D$ be a recurrent SCC, let $C \subseteq D$ be strongly
  connected, and let $c = \< q s \> \in C$.  Define $A:=B_{C,C}$ and
  write
\[ E^C_c := \{ w \in \Paths_s^\omega(M) : \exists\, v \in \Paths_c^\omega(A) \mbox{ s.t. } \pi_S(v)=w \} \, . \]
Then $\Pr_s(E^C_c)>0$ iff $C=D$.
\label{prop:recurrent}
\end{proposition}
\begin{proof}
Let $\vec{x}\geqslant \vec{0}$ be a dominant eigenvector of $A$ and write
$x_{\min}$ and $x_{\max}$ for the
respective minimum and maximum entries of $\vec{x}$.  Since $C$
is strongly connected, by Theorem~\ref{thm:irreducible}(1) and~(2), we have $x_{\min}>0$.

Intuitively, $E^C_c$ contains those paths in~$M$ that correspond to a path in~$A$.
For $n \in \Nat$ and $d \in C$, recall the notation $E^{C,n}_{c,d}$  
from Proposition~\ref{prop:powers-of-B}.
Write also $E^{C,n}_{c}:=\bigcup_{d \in C} E^{C,n}_{c,d}$.
Intuitively, $E^{C,n}_{c}$ contains those paths in~$M$ that correspond to a path in~$A$ for at least $n$ steps.
Then $E^C_c = \bigcap_{n\in\mathbb{N}} E^{C,n}_{c}$ by K\H{o}nig's Lemma.
We have for any $d \in C$:
\begin{equation}
\begin{aligned}
(A^n)_{c,d} \mathop{=}^\text{Prop.~\ref{prop:powers-of-B}} \Pr_s(E^{C,n}_{c,d}) \leqslant \Pr_s(E^{C,n}_{c})
\leqslant \sum_{d' \in C} \Pr_s(E^{C,n}_{c,d'}) = \sum_{d' \in C} (A^n)_{c,d'}
\end{aligned}
\label{eq:sandwich}
\end{equation}

  Suppose now that $C=D$. Then $\rho(A)=1$ and thus
\begin{align*}
|C| \cdot \Pr_s(E^{C,n}_{c}) & \ \geqslant \   \sum_{d\in C} (A^n)_{c,d} && \text{by Equation~\eqref{eq:sandwich}}\\
& \ = \ (A^n \vec{1})_c && \\
& \ \geqslant \ \textstyle\frac{1}{x_{\max}} (A^n \vec{x})_c
&& \text{since
$\vec{x} \leqslant x_{\max} \cdot \vec{1}$}\\
& \ = \ \textstyle\frac{\vec{x}_c}{x_{\max}} && \text{since
$A\vec{x}=\vec{x}$}
\end{align*}
By continuity of measures it follows that $\Pr_s(E^C_c)>0$.

Conversely, suppose that $C\neq D$.  Then
\begin{align*}
\Pr_s(E^{C,n}_{c}) & \ \leqslant \ \sum_{d \in C} (A^n)_{c,d}&& \text{by Equation~\eqref{eq:sandwich}}\\
& \ = \ (A^n \vec{1})_c && \\
& \ \leqslant \ \textstyle\frac{1}{x_{\min}} (A^n \vec{x})_c &&
\text{since $x_{\min} \cdot \vec{1} \leqslant \vec{x}$} \\
& \ = \ \rho(A)^n \textstyle \frac{\vec{x}_c}{x_{\min}} &&
\text{since $A\vec{x}=\rho(A)\vec{x}$}\end{align*}
But $\rho(A) < 1$ by Theorem~\ref{thm:irreducible}(3).
By continuity of measures it follows that $\Pr_s(E^C_c)=0$.  \qed
\end{proof}

In the following proposition, $E^{<D}$ includes those infinite paths in~$M$ that correspond to paths in~$B$ that eventually linger in~$B_{C,C}$ for a strict subset~$C$ of an SCC~$D$ in~$B$.
By Proposition~\ref{prop:recurrent} such paths have measure zero.
\begin{proposition}
Let $D$ be a recurrent SCC.
Write
\[ E^{<D} := \bigcup_{u \in S^*} \bigcup_{C \subsetneq D} \bigcup_{d \in C} u E^C_d \, . \]
Then $\Pr_s(E^{<D})=0$ for all $s \in S$.
\label{prop:recurrent-2}
\end{proposition}
\begin{proof}
Let $s \in S$, and $u \in S^*$, and $C \subsetneq D$ be strongly connected, and $\<r t\> = d \in C$.
Then $\Pr_s(u E^C_d) \leqslant \Pr_t(E^C_d) = 0$ by Proposition~\ref{prop:recurrent}.
Thus $E^{<D}$ is a countable union of sets of measure zero with respect to~$\Pr_s$.
It follows $\Pr_s(E^{<D})=0$.
\qed
\end{proof}

\paragraph{Proof of Lemma~\ref{lem:recurrent}(2)}
Let $d = \< q s \> \in D$.
We show that $\vec{z}_d > 0$ iff $D$ is accepting.

Suppose that $D$ is accepting.
Recall the definitions of $E_d^C$ and~$E^{<D}$ in Propositions \ref{prop:recurrent} and~\ref{prop:recurrent-2}.
Since $D$ is accepting, we have $E_d^D \setminus E^{<D} \subseteq \cL_q$.
Hence:
\begin{align*}
\vec{z}_d
& \ = \ \Pr_s(\cL_q) \\
& \ \geqslant \ \Pr_s(E_d^D \setminus E^{<D}) &&\\
& \ \geqslant \ \Pr_s(E_d^D) - \Pr_s(E^{<D}) && \\
  & \ > \ 0  && \text{by Propositions \ref{prop:recurrent} and~\ref{prop:recurrent-2}}
\end{align*}

Suppose now that $D$ is not accepting.  Let $v \in \Paths_d^\omega(B)$
be such that $\inf(\pi_Q(v)) \in \Acc$.  Since $D$ is not accepting,
$v$ must eventually either remain in a strongly connected set $C
\subsetneq D$ or reach an SCC $C \prec D$.  Thus we have
\begin{gather}
\Paths_s^\omega(M)\cap \cL_q \ \subseteq \ E^{<D} \cup
\bigcup_{u\in S^*} \bigcup_{C \prec D}\bigcup_{\<r t\> \in C}
u ( \Paths_t^\omega(M) \cap \cL_r ) \,.
\label{eq:meas-zero}
\end{gather}
We have $\Pr_s(E^{<D}) = 0$ by Proposition~\ref{prop:recurrent-2}.
For any SCC $C \prec D$, $\<r t\> \in C$, and $u\in S^*$, we
have
\[ \Pr_s(u ( \Paths_t^\omega(M) \cap \cL_r ) ) \ \leqslant \ \Pr_t( \Paths_t^\omega(M) \cap \cL_r ) \ = \
  \vec{z}_{\<r t\>} \ = \ 0 \, , \] with the last equality following from
Proposition~\ref{prop:lower-matrices-zero}.  Thus the right-hand side
of \eqref{eq:meas-zero} is a countable union of sets of measure zero
with respect to $\Pr_s$.  It follows that
$\vec{z}_{\<q s\>} = \Pr_s(\Paths_s^\omega(M)\cap \cL_q)=0$.
\qed

\section{Proof of Lemma~\ref{lem:cut}} 
\label{sec:cut-proof}
In this section we prove Lemma~\ref{lem:cut}, which is restated here.

\lemcut*

\subsection{Proof of Lemma~\ref{lem:cut}(1)}
\paragraph{Proof of Lemma~\ref{lem:cut}(1)}
Let $D \subseteq Q \times S$ be an SCC.
We have for all $\<q s\>, \<r t\> \in D$:
\begin{equation} \label{eq:An}
\begin{aligned}
\Pr_{s}\big(E^{D,n}_{\<q s\>, \<r t\>}\big) 
&\ = \ \Pr_{s}(\{s_0 s_1 s_2 \cdots \in \Paths^\omega_{s}(M): \<r t\> \in \<q s\> \then s_1 \cdots s_n\})
\end{aligned}
\end{equation}
Define $A := B_{D,D}$.
By Theorem~\ref{thm:irreducible}(1), $A$ has a dominant
eigenvector~$\vec{x}$, positive in all entries, with $A \vec{x} =
\rho(A) \vec{x}$.  Write $x_{\min} > 0$ for the smallest entry
of~$\vec{x}$.  For any $d_0 = \<q_0 s_0\> \in D$ and $n \in \Nat$
define a random variable $X^{d_0}_n : \Paths^\omega_{s_0}(M) \to
\Real_{\geqslant 0}$ by
\[
X^{d_0}_n(s_0 s_1 \cdots) \ = \ \sum_{d \,\in\, d_0 \then s_1 \cdots s_n} \vec{x}_{d}\,.
\]
We have:
\begin{equation} \label{eq:Ex-X}
\Ex_{s_0}(X^{d_0}_n) \ \mathop{=}^{\eqref{eq:An}} \ \sum_{d \in D} \Pr_{s_0}\big(E^{D,n}_{d_0,d}\big) \cdot \vec{x}_d \ \mathop{=}^\text{Prop.~\ref{prop:powers-of-B}} \ (A^n \vec{x})_{d_0} \ = \ \rho(A)^n \vec{x}_{d_0}\,,
\end{equation}
where $\Ex_{s_0}$ denotes expectation with respect to~$\Pr_{s_0}$.

Towards the direction~``$\Longleftarrow$'', let $d_0 \then s_1 \cdots s_k$ be a cut, where $d_0 = \<q_0 s_0\>$.
Then for all $w \in \Paths_{s_k}(M)$ we have $d_0 \then s_1 \cdots s_k w \ne \emptyset$.
So we have for all $n \geqslant k$, using Markov's inequality:
\begin{align*}
0 \ < \ \Pr_{s_0}(s_0 \cdots s_k S^\omega) \cdot x_{\min} &\ \leqslant \ \Pr_{s_0}(X^{d_0}_n \geqslant x_{\min}) \cdot x_{\min} \\
&\ \leqslant \ \Ex_{s_0}(X^{d_0}_n) \ \mathop{=}^{\eqref{eq:Ex-X}} \ \rho(A)^n \vec{x}_{d_0}
\end{align*}
This gives a uniform lower bound on~$\rho(A)^n$ for all $n \geqslant k$.
Hence $\rho(A) \geqslant 1$, and so $D$ is recurrent.

Towards the converse~``$\Longrightarrow$'', suppose that $D$ has no cuts.
Let $d_0 = \<q_0 s_0\> \in D$.
Since $D$ has no cuts, for any set $d_0 \then v \subseteq D$ there exists $w \in S^*$ with $d_0 \then v w = \emptyset$.
It follows that there are $\ell \in \Nat$ and $y > 0$ such that for all $s_0 \cdots s_n \in \Paths_{s_0}(M)$:
\[
\Pr_{s_0}(0 = X^{d_0}_{n+\ell} = X^{d_0}_{n+\ell+1} = \ldots \mid s_0 \cdots s_n S^\omega) \ \geqslant \ y
\]
Thus, for all $m \geqslant 0$:
\[
\Pr_{s_0}(0 = X^{d_0}_{\ell m} = X^{d_0}_{\ell m + 1} = \ldots) \ \geqslant \ 1 - (1-y)^m
\]
Hence, $X^{d_0}_0, X^{d_0}_1, \ldots$ converges to~$0$ almost surely with respect to~$\Pr_{s_0}$.
Since $X^{d_0}_0, X^{d_0}_1, \ldots$ is bounded (by $\sum_{d \in D} \vec{x}_d$), it follows $\lim_{n \to \infty} \Ex_{s_0}(X^{d_0}_n) = 0$.
By~\eqref{eq:Ex-X}, we conclude $\rho(A) < 1$, i.e., $D$ is not recurrent.
\qed

\subsection{Proof of Lemma~\ref{lem:cut}(2)}
For the proof of Lemma~\ref{lem:cut}(2)  we use the following standard
fact about model checking Markov chains:
\begin{lemma} \label{lem:regular}
Let $\cM = (S,M)$ be a Markov chain, and $\cL \subseteq S^\omega$ an $\omega$-regular language.
Suppose $s_0 \in S$ such that $\Pr_{s_0}(\cL) < 1$.
Then there exists $s_0 \cdots s_n \in \Paths_{s_0}(M)$ such that $\Pr_{s_n}(\{ w \in s_n S^\omega : s_0 \cdots s_{n-1} w \in \cL \}) = 0$.
\end{lemma}
\begin{proof}[sketch]
Let $\cA$ be a deterministic Muller automaton that accepts~$\cL$.
Let $q_0$ be the initial state of~$\cA$.
Since $\Pr_{s_0}(\cL) < 1$, there is a path $\<q_0 s_0\> \cdots \<q_n s_n\>$ in the product of $\cA$ and~$\cM$ that leads to a bottom SCC that is non-accepting, i.e., whose corresponding set of automaton states does not satisfy the Muller acceptance condition.
Then $s_0 \cdots s_n$ has the required properties.
\qed
\end{proof}

\paragraph{Proof of Lemma~\ref{lem:cut}(2)}

Let $D \ni \<q_0 s_0\>$ be a recurrent SCC, and let $\alpha = \beta \times \{s_n\} = \<q_0 s_0\> \then s_1 \cdots s_n \subseteq D$ be a cut.
Recall that $\beta \subseteq \delta(q_0, s_0 \cdots s_{n-1})$ and define $\beta' := \delta(q_0, s_0 \cdots s_{n-1}) \setminus \beta$.
Since all elements of $\delta(q_0, s_0 \cdots s_{n-1}) \times \{s_n\}$ are reachable from~$\<q_0 s_0\>$ in~$B$, by Proposition~\ref{prop:lower-matrices-zero} we have $\vec{z}_{\beta' \times \{s_n\}} = \vec{0}$, thus $\sum_{q \in \beta'} \Pr_{s_n}(\cL_q) = 0$.
Let $\vec{\mu} \in \{0,1\}^D$ be the cut vector associated with~$\alpha$.
Then we have:
\begin{align*}
\vec{\mu}^\top \vec{z}_D
&\ = \ \sum_{d \in \alpha} \vec{z}_d  \ = \ \sum_{q \in \beta} \Pr_{s_n}(\cL_q) \ = \ \sum_{q \in \beta} \Pr_{s_n}(\cL_q) + \sum_{q \in \beta'} \Pr_{s_n}(\cL_q) \\
&\ = \ \sum_{q \in \delta(q_0, s_0 \cdots s_{n-1})} \Pr_{s_n}(\cL_q) 
\ = \ \Pr_{s_n}( \cL(\cA[\delta(q_0, s_0 \cdots s_{n-1})]))\,, 
\end{align*}
where the last equality holds as the sets $\cL_q$ are disjoint by unambiguousness.
Hence $\vec{\mu}^\top \vec{z}_D \leqslant 1$.
%
%
%
%
Suppose $\vec{\mu}^\top \vec{z}_D < 1$.
Then $\Pr_{s_n}( \cL(\cA[\delta(q_0, s_0 \cdots s_{n-1})])) < 1$.
Then by Lemma~\ref{lem:regular} there exists $s_n \cdots s_m \in \Paths_{s_n}(M)$ such that
\[
\Pr_{s_m}(\{ w \in s_m S^\omega \, : \, s_n \cdots s_{m-1} w \in \cL(\cA[\delta(q_0, s_0 \cdots s_{n-1})]) \}) \ = \ 0\,.
\]
Equivalently,
\begin{equation} \label{eq:cut=normalizer-proof}
\Pr_{s_m}(\{ w \in s_m S^\omega \, : \, w \in \cL(\cA[\delta(q_0, s_0 \cdots s_{m-1})]) \}) \ = \ 0\,.
\end{equation}
But since $\alpha$ is a cut, we have $\<q_0 s_0\> \then s_1 \cdots s_m \ne \emptyset$, i.e, there exists $q \in \delta(q_0, s_0 \cdots s_{m-1})$ with $\<q s_m\> \in D$.
By~\eqref{eq:cut=normalizer-proof} we have $\Pr_{s_m}(\cL_q) = 0$.
With Lemma~\ref{lem:recurrent}(2) it follows that $D$ is not accepting.
\qed

\subsection{Proof of Lemma~\ref{lem:cut}(3)} \label{sub:computing-cuts}

Let $D \subseteq Q \times S$ be a recurrent SCC.
In this section we show how to compute a cut in polynomial time.
By Lemma~\ref{lem:cut}(2), this also yields a $D$-normalizer if $D$ is accepting.

Since automaton $\cA$ is diamond-free, we have that if $d \then v \supseteq \{d_1, d_2\}$ with $d_1 \ne d_2$ then any sets $d_1 \then w$ and~$d_2 \then w$ are disjoint.
\begin{lemma} \label{lem:get-larger}
Let $D \subseteq Q \times S$ be a recurrent SCC.
Let $d \in D$.
Suppose $w \in S^*$ is such that $d \then w \ni d$ is not a cut.
Then there are $v \in S^*$ and $d' \ne d$ with $d \then v \supseteq \{d,d'\}$ and $d' \then w \ne \emptyset$.
Hence $d \then v w \supseteq \{d,d'\} \then w  = d \then w \cup d' \then w \supsetneq d \then w$.
\end{lemma}
\begin{proof}
Since $D$ is recurrent, by Lemma~\ref{lem:cut}(1), $D$ has a cut~$\alpha$, say $\alpha = d_1 \then v_1$.
Since $D$ is an SCC, there is $v_0$ with $d \then v_0 \ni d_1$, hence $d \then v_0 v_1 \supseteq \alpha$ is also a cut.
Again, since $D$ is an SCC, there is $v_2$ with $d \then v_0 v_1 v_2 \ni d$.
Define $v := v_0 v_1 v_2$.
Then $d \then v \ni d$ is a cut.
Moreover, we have $d \then v w \supseteq d \then w$.
Since $d \then v w$ is a cut but $d \then w$ is not, we have:
\[
 d \then w \cup (d \then v \setminus \{d\}) \then w \ = \ d \then v w \ \supsetneq \ d \then w
\]
So there is $d' \in d \then v \setminus \{d\}$ with $d' \then w \ne \emptyset$.
\qed
\end{proof}

\begin{lemma} \label{lem:compute-cut}
Let $D \subseteq Q \times S$ be a recurrent SCC.
Let $d = \<q_0 s_0\> \in D$.
The following algorithm is a polynomial-time algorithm that computes $w \in S^*$ with $|w| \leqslant |Q|^3 |S|$ such that $d \then w$ is a cut of~$D$:
\begin{enumerate}
\item $w := \varepsilon$ (the empty word)
\item while $\exists\, v \in S^*$ and $\exists\, d' \ne d$ such that $d \then v \supseteq \{d, d'\}$ and $d' \then w \ne \emptyset:$ \\
\mbox{}\hspace{4mm} $w := v w$
\item return $d \then w$
\end{enumerate}
\end{lemma}
\begin{proof}
By Lemma~\ref{lem:get-larger} the algorithm returns a cut.
In every iteration, the set~$d \then w$ increases (cf.~Lemma~\ref{lem:get-larger}), so the algorithm terminates after at most $|Q|$ iterations.

Consider the directed graph~$G = (V,E)$ with
\begin{align*}
V &\ = \ \{ (q, q', s) \in Q \times Q \times S \, : \, \<q s\>, \<q' s\> \in D \} \\
E &\ = \ \{ (q, q', s) \to (r, r', t) \, : \, \delta(q,s) \ni r,\ \delta(q',s) \ni r',\ M_{s,t} > 0\}\,.
\end{align*}
Then for any $v = s_1 \cdots s_n \in S^*$ and $d_n = \<q_n s_n\>, d_n' = \<q_n' s_n\> \in D$ we have $d \then v \supseteq \{d_n, d_n'\}$ if and only if there are $q_1, q_1', \ldots, q_{n-1}, q_{n-1}'$ such that
\[
(q_0,q_0,s_0) \to (q_1,q_1',s_1) \to \cdots \to (q_n,q_n',s_n)
\]
is a path in~$G$.
It follows that with a (polynomial-time) reachability analysis of~$G$ one can compute all $d' \in D$ for which there exists $v_{d'} \in S^*$ with $d \then v_{d'} \supseteq \{d, d'\}$.
The shortest such~$v_{d'}$ correspond to shortest paths in~$G$, hence satisfy $|v_{d'}| \leqslant |V| \leqslant |Q|^2 |S|$.
Moreover, one can check in polynomial time whether $d' \then w \ne \emptyset$.
\qed
\end{proof}

\section{An NC Model Checking Procedure} \label{sec:NC}

In this section we show that one can model check Markov chains against unambiguous automata in NC.
To achieve this we strengthen our assumptions on the acceptance condition:
we assume that the given automaton $\cA = (Q,\Sigma,\delta,Q_0,\Acc)$ has an NC-decidable acceptance condition; i.e., given~$\cA$ and a set $R \subseteq Q$, one can compute in~NC whether $R \in \Acc$.
This is the case, e.g., if $\Acc$ is given as a B\"uchi condition.
We show:
\begin{theorem} \label{thm:NC}
  Given a Markov chain $\cM$, an initial distribution~$\iota$, and an unambiguous automaton~$\cA$ with NC-decidable acceptance condition,
  the value $\Pr_\iota(\cL(\cA))$ is computable in~NC.
\end{theorem}
As a consequence, our approach yields optimal complexity for model checking Markov chains against LTL specifications:
\begin{corollary} \label{cor:LTL-PSPACE} Given a Markov chain $\cM$,
  an initial distribution~$\iota$, and an LTL formula~$\varphi$, the
  value $\Pr_\iota(\cL(\varphi))$ is computable in PSPACE.
\end{corollary}
\begin{proof}
  There is a classical polynomial-space procedure that
  translates~$\varphi$ into an (exponential-sized) B\"{u}chi automaton
  $\cA_\varphi$~\cite{VardiWolper86}.  As noted by several authors
  (e.g.,~\cite{ChaKat14,CouSahSut03}), this procedure can easily be
  adapted to ensure that $\cA_\varphi$ be a UBA.

  Now recall from Section~\ref{sec:prelim} that a function that is
  computable in NC is also computable in poly-logarithmic space. By
  Theorem~\ref{thm:NC} it follows that we can compute
  $\Pr_\iota(\cL(\varphi))$ in poly-logarithmic space in $\cA_\varphi$
  and $\cM$.  Thus using standard techniques for composing
  space-bounded transducers (see, e.g.,~\cite[Proposition
  8.2]{Pap94}), we can compute $\Pr_\iota(\cL(\varphi))$ using
  polynomial space in $\varphi$ and $\cM$. \qed
\end{proof}

Towards a proof of Theorem~\ref{thm:NC}, observe that most steps of
the algorithm from the proof of Theorem~\ref{thm:PMC-MC-UBA} can be
implemented in~NC in a straightforward way.  The exception is the
cut-computation algorithm from Lemma~\ref{lem:compute-cut}, which
seems inherently sequential.  Recall that we used this algorithm
because a cut vector of an accepting recurrent SCC~$D$ yields a
$D$-normalizer (Lemma~\ref{lem:cut}(2)) and we need such a normalizer
to set up up the equation system~\eqref{eq:linear-system} from
Lemma~\ref{lem:linear-system}.  Note that any convex combination of
normalizers is also a normalizer.  In the following we show how to
compute such a normalizer in~NC.

Let $D \subseteq Q \times S$ be a recurrent SCC.
Let $d_0 = \<q_0 s_0\> \in D$.
Define $E := \{d \in D : \exists\, v \in S^*\,.\, d_0 \then v \supseteq \{d_0,d\}\}$.
Observe that $E$ is fibered on~$s_0$, i.e., there is $R \subseteq Q$ with $E = R \times \{s_0\}$.
For any $q \in Q$ and $w \in S^*$ and $\alpha \subseteq Q$ with $\<q s_0\> \then w = \alpha \times \{s_0\}$ we write $q \then w = \alpha$ to avoid clutter.
Hence
\[
 R \ = \ \{q \in Q \,:\, \exists\, v \in S^* .\, q_0 \then v \supseteq \{q_0, q\}\}\,.
\]
One can compute~$R$ in~NC by a graph reachability analysis.
Similarly, one can compute in~NC for any $q \in R$ a word $v_q \in S^*$ such that $q_0 \then v_q \supseteq \{q_0,q\}$.
One can also compute in~NC for any $q \in R$ a matrix $A(q) \in \{0,1\}^{R \times R}$ such that $A(q)_{r,r'} = 1$ if and only if $r \then v_q \ni r'$.
Define
\[
 A \ := \ \frac{1}{|R|} \sum_{q \in R} A(q) \,.
\]
In the following, for a set $\alpha \subseteq R$, we write $\vec{\chi}(\alpha) \in \{0,1\}^R$ for the characteristic vector of~$\alpha$. 
If $\alpha$ is a singleton set~$\{q\}$ we may write $\vec{\chi}(q)$ for~$\vec{\chi}(\alpha)$.
The following lemma provides a $D$-normalizer:
\begin{lemma} \label{lem:A-limit}
Let $D \subseteq Q \times S$ be an accepting recurrent SCC.
Let $d_0 = \<q_0 s_0\> \in D$.
Define $R \subseteq Q$ and~$A \in [0,1]^{R \times R}$ as above.
Then the limit
\[
 \vec{\eta}^\top \ := \ \lim_{n \to \infty} \vec{\chi}(q_0)^\top A^n \ \in \ [0,1]^R
\]
exists.
The vector $\vec{\mu} \in [0,1]^D$ with
\[
\vec{\mu}_{\<q s\>} =
\begin{cases}
    \vec{\eta}_{q} & \text{if } q \in R \text{ and } s = s_0 \\[0.5ex]
    0 & \text{otherwise}
\end{cases}
\]
is a $D$-normalizer.
\end{lemma}
\begin{proof}
Observe that $A(q)_{q_0,q_0} = A(q)_{q_0,q} = 1$ for all $q \in R$.
For any $q_1, \ldots, q_n \in R$ we have $q_0 \then v_{q_1} \cdots v_{q_n} \supseteq \{q_0,q_n\}$.
Since the automaton~$\cA$ is diamond-free, we have more generally:
\begin{equation} \label{eq:then=mult}
  \vec{\chi}(q_0)^\top A(q_1) \cdot \ldots \cdot A(q_n) \ = \ \vec{\chi}(q_0 \then v_{q_1} \cdots v_{q_n})^\top
\end{equation}
It follows that, for all~$n$, the entries of $\vec{\chi}(q_0)^\top A^n$ are at most~$1$, and the $q_0$-entry equals~$1$.
Since $A$ is nonnegative, we obtain:
\[
 \vec{\chi}(q_0)^\top \ \leqslant \ \vec{\chi}(q_0)^\top A \ \leqslant \ \vec{\chi}(q_0)^\top A^2 \ \leqslant \ \cdots
\]
So the limit $\vec{\eta}^\top \in [0,1]^R$ exists.

\newcommand{\Cuts}{\mathit{Cuts}}%
Define $\Cuts := \{\alpha \subseteq R : \alpha \times \{s_0\} \text{ is a cut}\}$ and $V := \{v_q \in S^* : q \in R\}$.
It follows from the definition of the~$v_q$ that we have $q_0 \then v \ni q_0$ for all $v \in V^*$.
By Lemma~\ref{lem:get-larger} there are $\bar q_1, \ldots, \bar q_k \in R$ such that
\[
q_0 \ = \ q_0 \then \varepsilon \ \subsetneq \ q_0 \then v_{\bar q_1} \subsetneq \ q_0 \then v_{\bar q_2} v_{\bar q_1} \ \subsetneq \ \ldots \ \subsetneq \ q_0 \then v_{\bar q_k} \cdots v_{\bar q_2} v_{\bar q_1}
\]
and $q_0 \then \bar v \in \Cuts$, where $\bar v = v_{\bar q_k} \cdots v_{\bar q_1}$.
Thus, if $w \in V^* \cdot \{ \bar v \} \cdot V^*$ then $q_0 \then w \in \Cuts$.

Define a function $\vec{\nu} : [0,1]^R \to [0,1]^D$ with
\[
\vec{\nu}(\vec{x})_{\<q s\>} =
\begin{cases}
    \vec{x}_{q} & \text{if } q \in R \text{ and } s = s_0 \\[0.5ex]
    0 & \text{otherwise.}
\end{cases}
\]
By Lemma~\ref{lem:cut}(2) for all $\alpha \in \Cuts$ the vector $\vec{\nu}(\vec{\chi}(\alpha))$ is a $D$-normalizer, i.e., $\vec{\nu}(\vec{\chi}(\alpha))^\top \vec{z}_D = 1$.
Defining $f : [0,1]^R \to [0,D]$ with $f(\vec{x}) = \vec{\nu}(\vec{x})^\top \vec{z}_D$, we have $f(\vec{\chi}(\alpha)) = 1$ for all $\alpha \in \Cuts$.
Note that $f$~is a linear function.

Consider the stochastic process $r_1, r_2, \ldots$ where the $r_i$ are chosen from~$R$ independently and uniformly at random.
Write $\Pr$ and~$\Ex$ for the associated probability measure and expectation.
It follows from~\eqref{eq:then=mult} that we have
\begin{equation} \label{eq:NC-Expectation}
 \vec{\chi}(q_0)^\top A^n \ = \ \Ex\left(\vec{\chi}(q_0 \then v_{r_1} \cdots v_{r_n})^\top\right) \quad \text{for all $n \in \Nat$}\,.
\end{equation}

In the following, when we say `almost surely' we mean with probability~$1$ with respect to~$\Pr$.
Almost surely, $\bar q_k, \ldots, \bar q_1$ will eventually appear as a contiguous subsequence of $r_1, r_2, \ldots$.
That is, $v_{r_1} v_{r_2} \cdots \in V^* \cdot \{\bar v\} \cdot V^\omega$ almost surely.
Thus, almost surely there is $m \in \Nat$ such that $q_0 \then v_{r_1} \cdots v_{r_n} \in \Cuts$ holds for all $n \geqslant m$.
Hence,
\begin{equation} \label{eq:A-limit-key}
\Pr\left(\lim_{n \to \infty} f\left(\vec{\chi}(q_0 \then v_{r_1} \cdots v_{r_n})\right) \ = \ 1\right) \ \ = \ \ 1\,.
\end{equation}
So we have
\begin{align*}
\vec{\mu}^\top \vec{z}_D \
&=\ \vec{\nu}(\vec{\eta})^\top \vec{z}_D && \text{definitions of $\vec{\mu}, \vec{\nu}$} \\
&=\ f(\vec{\eta}) && \text{definition of $f$} \\
&=\ f\Big(\Big(\lim_{n \to \infty} \vec{\chi}(q_0)^\top A^n\Big)^\top\Big) && \text{definition of~$\vec{\eta}$} \\
&=\ f\left(\lim_{n \to \infty} \Ex\left(\vec{\chi}(q_0 \then v_{r_1} \cdots v_{r_n})\right)\right) && \text{by \eqref{eq:NC-Expectation}} \\
&=\ \lim_{n \to \infty} f\left(\Ex\left(\vec{\chi}(q_0 \then v_{r_1} \cdots v_{r_n})\right)\right) && \text{$f$ is continuous} \\
&=\ \lim_{n \to \infty} \Ex\left(f\left(\vec{\chi}(q_0 \then v_{r_1} \cdots v_{r_n})\right)\right) && \text{$f$ is linear} \\
&=\ \Ex \lim_{n \to \infty} \left(f\left(\vec{\chi}(q_0 \then v_{r_1} \cdots v_{r_n})\right)\right) && \text{dominated convergence theorem} \\
&=\ 1 && \text{by~\eqref{eq:A-limit-key}\,;}
\end{align*}
i.e., $\vec{\mu}$ is a $D$-normalizer.
\qed
\end{proof}

\begin{lemma} \label{lem:xi}
The vector~$\vec{\eta}$ with $\vec{\eta}^\top = \lim_{n \to \infty} \vec{\chi}(q_0)^\top A^n$ from Lemma~\ref{lem:A-limit} is the unique solution of the linear system
\begin{equation} \label{eq:xi}
\vec{\xi}^\top \ = \ \vec{\xi}^\top (A - E) + \vec{\chi}(q_0)^\top
\end{equation}
where $\vec{\xi}$ is a vector of variables indexed by~$R$, and $E \in \{0,1\}^{R \times R}$ is the matrix with $E_{q,r} = 1$ if and only if $q=r=q_0$.
\end{lemma}
\begin{proof}
Recall from the proof of Lemma~\ref{lem:A-limit} that we have $\left(\vec{\chi}(q_0)^\top A^n\right)_{q_0} = 1$ for all~$n$.
It follows that $\vec{\eta}_{q_0} = 1$ and thus $\vec{\eta}^\top E = \vec{\chi}(q_0)^\top$.
Hence we have:
\[
\vec{\eta}^\top\ = \ \vec{\eta}^\top A  \ = \ \vec{\eta}^\top (A - E) + \vec{\eta}^\top E \ = \ \vec{\eta}^\top (A - E) + \vec{\chi}(q_0)^\top
\]
So $\vec{\eta}$ solves~\eqref{eq:xi}.
Towards uniqueness, since
\begin{align*}
\vec{\eta}^\top
& \ = \ \vec{\eta}^\top (A - E) + \vec{\chi}(q_0)^\top \\
& \ = \ \left( \vec{\eta}^\top (A-E) + \vec{\chi}(q_0)^\top \right) (A - E) + \vec{\chi}(q_0)^\top\,,
\end{align*}
we have:
\begin{align*}
\vec{\eta}^\top (A-E)^2
& \ = \ \vec{\eta}^\top - \vec{\chi}(q_0)^\top A  + \vec{\chi}(q_0)^\top E - \vec{\chi}(q_0)^\top \\
& \ = \ \vec{\eta}^\top - \vec{\chi}(q_0)^\top A
\end{align*}
Since $A_{q_0,q} > 0$ holds for all $q \in R$, it follows that $\vec{\eta}^\top (A-E)^2$ is less than~$\vec{\eta}^\top$ in all entries.
Hence $\rho(A-E) < 1$.
Let $\vec{x}$ be any solution of~\eqref{eq:xi}.
Then $\vec{x}^\top - \vec{\eta}^\top = \left(\vec{x}^\top - \vec{\eta}^\top\right) (A-E)$.
Since $\rho(A-E) < 1$, it follows $\vec{x} = \vec{\eta}$.
\qed
\end{proof}

Now we can prove Theorem~\ref{thm:NC}.
\begin{proof}[of Theorem~\ref{thm:NC}]
We follow the same approach as in the proof of Theorem~\ref{thm:PMC-MC-UBA}.
Most steps can easily be carried out in~NC.
Instead of step~4, we compute, in~NC \cite[Theorem~5]{BorodinGathenHopcroft82}, the vector~$\vec{\eta}$ by solving the linear system~\eqref{eq:xi} in Lemma~\ref{lem:xi}.
From~$\vec{\eta}$ we easily obtain a $D$-normalizer by Lemma~\ref{lem:A-limit}.
\qed
\end{proof}


\section{Implementation and Experiments}
\label{uba_implementation}


We implemented a probabilistic model checking procedure for Markov chains
and UBA specifications using the algorithm detailed in Section~\ref{sec:uba}
as an extension to the probabilistic model checker
\prism{}~\cite{prism40} version 4.4 beta.
\footnote{%
    More details are available
    at~\url{https://wwwtcs.inf.tu-dresden.de/ALGI/TR/JCSS19/}.
}
 All experiments were carried out on a computer with
 two
 Intel E5-2680 8-core CPUs at 2.70~GHz with \(384\,\mathrm{GB}\) of RAM running Linux, a time
 limit of \(30\) minutes and a memory limit of \(10\,\mathrm{GB}\).
%
Our implementation is based on the \explicitengine{} engine of \prism{}, where
the Markov chain is represented explicitly.
%
%
Our implementation supports UBA-based model checking for handling the
LTL fragment of \prism's $\mathrm{PCTL}^*$-like specification language
as well as direct verification against a path specification given by a
UBA provided in the HOA format~\cite{Hanoi-CAV15}. For LTL formulas,
we rely on external LTL-to-UBA translators.
For the purpose of the benchmarks
we employ the \ltltotgba{} tool from \spot{}~\cite{Duret14} version 2.7
to generate a UBA for a given LTL formula.
%
For the linear algebra parts of the algorithms, we use the \colt{}
library~\cite{Hoschek04}.

\subsection{Analyzing SCCs}
In our experiments we solved the linear system~\eqref{eq:linear-system} SCC-wise, bottom-up.
We call accepting recurrent SCCs \emph{positive}.
We considered two different variants for checking positivity of an SCC.
The first variant relies on \colt{} to perform a QR
decomposition of the matrix for the SCC to compute the rank, which allows for
deciding the positivity of the SCC.
The second approach
is based on a variant of the power iteration
method for iteratively computing an eigenvector.

\paragraph{Recurrence check via rank computation}
\label{sec:rank}%

Let $D \subseteq Q \times S$ be an SCC.
Using Theorem~\ref{thm:nonnegative}(1) we can check whether $D$ is recurrent by seeing if the linear system $B_{D,D}\vec{x}=\vec{x}$ has a nonzero solution.
It is equivalent to check whether the matrix $B_{D,D} - I$ has full rank, where $I$ is the $D \times D$ identity matrix.
Indeed, if $B_{D,D} - I$ has full rank, then it has a trivial kernel $\{0\}^D$, so $B_{D,D}\vec{x}=\vec{x}$ does not have a nonzero solution.
Conversely, if $B_{D,D} - I$ does not have full rank, then there is a nonzero $\vec{x}$ with $B_{D,D}\vec{x}=\vec{x}$.

\paragraph{Iterative algorithm}

Consider again an SCC $D \subseteq Q \times S$, and define $\tB = (I + B_{D,D})/2$.
Denote by $\vone \in \{1\}^D$ the column vector all whose components are~$1$.
For $i \geqslant 0$ define $\vec{y}(i) = \tB^i \vone$.
Our algorithm is as follows.
Exploiting the recurrence $\vec{y}(i+1) = \tB \vec{y}(i)$ compute the sequence
$\vec{y}(0), \vec{y}(1), \ldots$ until we find $i > 0$ with either $\vec{y}(i+1)
< \vec{y}(i)$ (by this inequality we mean strict inequality in all components)
or $\vec{y}(i+1) = \vec{y}(i)$.
In the first case we conclude that $D$ is not recurrent. 

In the second case we conclude that $D$ is recurrent.
If $D$ is, in addition, accepting, then we can use the result of our iterative computation to simplify the linear system~\eqref{eq:linear-system}:
we compute a cut vector $\vec{\mu}$ and a scalar $c > 0$ so that $c \vec{\mu}^\top \vec{y}(i) = 1$, and replace all equations in~\eqref{eq:linear-system} with variables from $\vec{\zeta}_D$ on the left-hand side by $\vec{\zeta}_D = c \vec{y}(i)$.
This algorithm is justified by the following two lemmas, combined with Lemma~\ref{lem:cut}(2).

\begin{lemma} \label{lem-iter-zero-new}
$D$ is recurrent if and only if there is no $i \geqslant 0$ with
    $\vec{y}(i+1) < \vec{y}(i)$.
\end{lemma}

\begin{lemma} \label{lem-iter-pos-new}
If $D$ is recurrent, then $\vec{y}(\infty) := \lim_{i \to \infty}
    \vec{y}(i) > \vzero$ exists, and $B \vec{y}(\infty) = \vec{y}(\infty)$, and
    $\vec{z}_D$ is a scalar multiple of $\vec{y}(\infty)$.
\end{lemma}

For the proofs we need the following two auxiliary lemmas:
\begin{lemma} \label{lem-iter-aux-new}
    For any $\vec{y} \in \mathbb{C}^D$ and any $c \in \mathbb{C}$ we have $B_{D,D}
    \vec{y} = c \vec{y}$ if
    and only if $\tB \vec{y} = \frac{1+c}{2} \vec{y}$.
    In particular, $B_{D,D}$ and~$\tB$ have the same eigenvectors with eigenvalue~$1$.
\end{lemma}
\begin{proof}
Immediate.
\qed
\end{proof}

\begin{lemma} \label{lem-iter-matrix-conv-new}
Let $\rho > 0$ denote the spectral radius of~$\tB$.
Then the matrix limit $\lim_{i \to \infty} \left( \tB / \rho \right)^i$ exists and is strictly positive in all entries.
\end{lemma}
\begin{proof}
Since $B_{D,D}$ is irreducible, $\tB^{|D|}$ is strictly positive (in all entries).
By Theorem~\ref{thm:irreducible}(5) the matrix limit
 \[
  \lim_{i \to \infty} \left(\tB/\rho\right)^i = \lim_{i \to \infty}
  \left(\left(\tB/\rho\right)^{|D|}\right)^i
 \]
 exists and is strictly positive.
\qed
\end{proof}

\begin{proof}[of Lemma~\ref{lem-iter-zero-new}]
Let $i \geqslant 0$ with $\vec{y}(i+1) < \vec{y}(i)$.
It follows from Theorem~\ref{thm:irreducible}(4) that the spectral radius of~$\tB$ is $<1$.
Hence by Lemma~\ref{lem-iter-aux-new} the spectral radius of~$B_{D,D}$ is also $<1$, i.e., $D$ is not recurrent.

For the converse, suppose $D$ is not recurrent, i.e., the spectral radius of~$B_{D,D}$ is $<1$.
Let $\rho$ denote the spectral radius of~$\tB$.
By Lemma~\ref{lem-iter-aux-new}, we have $\rho < 1$.
If $\rho = 0$ then $\tB$ is the zero matrix and we have $\vec{y}(1) = \vzero <
\vone = \vec{y}(0)$.
Let $\rho > 0$.
It follows from Lemma~\ref{lem-iter-matrix-conv-new} that there is $i \geqslant 0$ such
that $\rho \left(\tB/\rho\right)^{i+1} < \left(\tB/\rho\right)^i$ (with the inequality strict in all components).
Hence $\tB^{i+1} < \tB^i$ and $\vec{y}(i+1) = \tB^{i+1} \vone < \tB^i \vone =
\vec{y}(i)$.
\qed
\end{proof}

\begin{proof}[of Lemma~\ref{lem-iter-pos-new}]
Let $D$ be recurrent, i.e., the spectral radius of~$B_{D,D}$ is~$1$.
So, with Lemma~\ref{lem-iter-aux-new} the spectral radius of~$\tB$ is~$1$.
By Lemma~\ref{lem-iter-matrix-conv-new} the limit $\vec{y}(\infty) = \lim_{i \to \infty}
\tB^i \vone$ exists and is positive.
From the definition of $\vec{y}(\infty)$ we have $\vec{y}(\infty) = \tB \vec{y}(\infty)$.
By Lemma~\ref{lem-iter-aux-new} also $\vec{y}(\infty) = B_{D,D} \vec{y}(\infty)$.
Lemma~\ref{lem:recurrent}(1) states $\vec{z}_D = B_{D,D} \vec{z}_D$.
By Theorem~\ref{thm:irreducible}(2), the eigenspace
of~$B_{D,D}$ associated with the spectral radius is one-dimensional, implying that $\vec{z}_D$ is a scalar multiple of~$\vec{y}(\infty)$.
\qed
\end{proof}

In our implementation we replace the check whether $\vec{y}(i+1) = \vec{y}(i)$ by a check whether $\vec{y}(i+1)$ and~$\vec{y}(i)$ are approximately equal, up to a convergence threshold of $10^{-10}$.
Thus, our implementation is sound only up to these numerical issues.

\subsection{Evaluation of the rank computation}

\label{sec:appendix-experiments}
\label{appendix:experiments}

The iterative (power iteration) algorithm from the previous subsection has the benefit that, in addition to deciding positivity of an SCC,
the computed eigenvector can be directly used to compute
the values of $\vec{z}$ corresponding to a positive SCC, once a cut has been found.
We have evaluated the performance and scalability of
the cut generation algorithm together with both approaches for
treating SCCs with selected automata specifications
that are challenging for our UBA-based model checking approach.

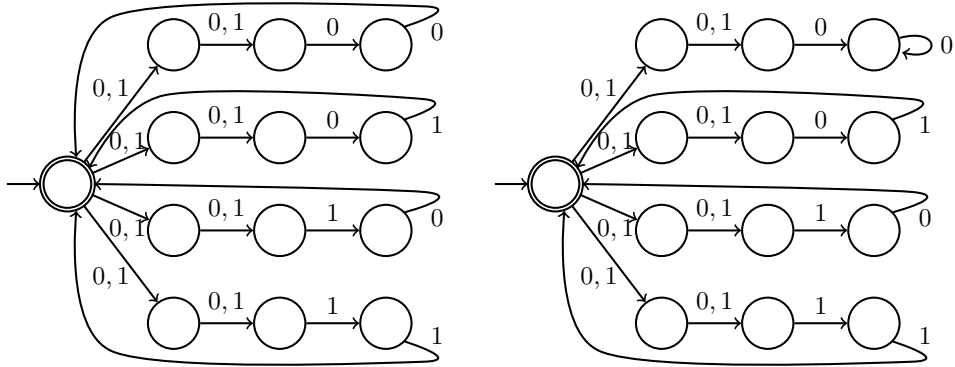
\begin{figure}[t]
    \hspace*{-0.97 em}
    \scalebox{0.97}{%
    \begin{tikzpicture}[initial text=,initial where=left, every node/.style={thick}, every path/.style={thick}]
        \node[state, initial, accepting, ellipse, minimum size=2 em, inner sep=0 em]    (q)     at(0,0) {};
        \coordinate                 (coord)             at(0,-0.2625 \textwidth) {};

        \foreach \line in {1,...,4}
        {
            \foreach \column in {1,...,3}
            {
                \node[state, ellipse, minimum size=2 em, inner sep=0 em] (q\column\line) at ($(coord) + (0.12 * \column * \textwidth, 0.105 * \line * \textwidth)$) {};
            }
            \draw[->]   (q1\line) edge[]  node[above] {\(0,1\)} (q2\line);
        }
        \draw[->]   (q)     edge[]  node[left, pos=0.75]  {\(0,1\)}   (q11);
        \draw[->]   (q)     edge[]  node[below, pos=0.62] {\(0,1\)}    (q12);
        \draw[->]   (q)     edge[]  node[above, pos=0.62, yshift=-0.5 ex] {\(0,1\)}    (q13);
        \draw[->]   (q)     edge[]  node[left, pos=0.75]  {\(0,1\)}   (q14);
        \draw[->]   (q21)   edge[]  node[above] {\(1\)}     (q31);
        \draw[->]   (q22)   edge[]  node[above] {\(1\)}     (q32);
        \draw[->]   (q23)   edge[]  node[above] {\(0\)}     (q33);
        \draw[->]   (q24)   edge[]  node[above] {\(0\)}     (q34);

        \draw[->] plot [smooth] coordinates {($(q34) + (0.7 em, 0.7 em)$) ($(q34) + (1.5 em, 1.5 em)$) ($(0.05 \textwidth, 0.16 \textwidth) + (0,1 em)$) (0.35 em, 1.05 em)};
        \node[draw=none]    (symb0) at($(q34) + (2 em, 0.5 em)$)  {\(0\)};
        \draw[->] plot [smooth] coordinates {($(q33) + (0.7 em, 0.7 em)$) ($(q33) + (1.5 em, 1.5 em)$) ($(0.11 \textwidth, 0.057 \textwidth) + (0, 1.5 em)$) (0.85 em, 0.62 em)};
        \node[draw=none]    (symb2) at($(q33) + (2 em, 0.5 em)$)  {\(1\)};
        \draw[->] plot [smooth] coordinates {($(q32) + (0.7 em, 0.7 em)$) ($(q32) + (1.5 em, 1.5 em)$)  (1.05 em, 0 em)};
        \node[draw=none]    (symb3) at($(q32) + (2 em, 0.5 em)$)  {\(0\)};
        \draw[->] plot [smooth] coordinates {($(q31) + (0.7 em, -0.7 em)$) ($(q31) + (1.5 em, -1.5 em)$) ($(0.05 \textwidth, -0.16 \textwidth) + (0,-1 em)$) ($(0.35 em, -1.05 em)$)};
        \node[draw=none]    (symb1) at($(q31) + (2 em, -0.5 em)$)  {\(1\)};
    \end{tikzpicture}%
    \begin{tikzpicture}[initial text=,initial where=left, every node/.style={thick}, every path/.style={thick}]
        \node[state, initial, accepting, ellipse, minimum size=2 em, inner sep=0 em]    (q)     at(0,0) {};
        \coordinate                 (coord)             at(0,-0.2625 \textwidth) {};

        \foreach \line in {1,...,4}
        {
            \foreach \column in {1,...,3}
            {
                \node[state, ellipse, minimum size=2 em, inner sep=0 em] (q\column\line) at ($(coord) + (0.12 * \column * \textwidth, 0.105 * \line * \textwidth)$) {};
            }
            \draw[->]   (q1\line) edge[]  node[above] {\(0,1\)} (q2\line);
        }
        \draw[->]   (q)     edge[]  node[left, pos=0.75]  {\(0,1\)}   (q11);
        \draw[->]   (q)     edge[]  node[below, pos=0.62] {\(0,1\)}    (q12);
        \draw[->]   (q)     edge[]  node[above, pos=0.62, yshift=-0.5 ex] {\(0,1\)}    (q13);
        \draw[->]   (q)     edge[]  node[left, pos=0.75]  {\(0,1\)}   (q14);
        \draw[->]   (q21)   edge[]  node[above] {\(1\)}     (q31);
        \draw[->]   (q22)   edge[]  node[above] {\(1\)}     (q32);
        \draw[->]   (q23)   edge[]  node[above] {\(0\)}     (q33);
        \draw[->]   (q24)   edge[]  node[above] {\(0\)}     (q34);

        \draw[->]       (q34)   edge[loop right]    node[right] {\(0\)} (q34);
        \draw[->] plot [smooth] coordinates {($(q33) + (0.7 em, 0.7 em)$) ($(q33) + (1.5 em, 1.5 em)$) ($(0.11 \textwidth, 0.057 \textwidth) + (0, 1.5 em)$) (0.85 em, 0.62 em)};
        \node[draw=none]    (symb2) at($(q33) + (2 em, 0.5 em)$)  {\(1\)};
        \draw[->] plot [smooth] coordinates {($(q32) + (0.7 em, 0.7 em)$) ($(q32) + (1.5 em, 1.5 em)$)  (1.05 em, 0 em)};
        \node[draw=none]    (symb3) at($(q32) + (2 em, 0.5 em)$)  {\(0\)};
        \draw[->] plot [smooth] coordinates {($(q31) + (0.7 em, -0.7 em)$) ($(q31) + (1.5 em, -1.5 em)$) ($(0.05 \textwidth, -0.16 \textwidth) + (0,-1 em)$) ($(0.35 em, -1.05 em)$)};
        \node[draw=none]    (symb1) at($(q31) + (2 em, -0.5 em)$)  {\(1\)};
    \end{tikzpicture}%
    }
\caption{UBA ``complete automaton'' (left) and ``nearly complete automaton''
(right) for $k=2$.}
\label{fig:uba-complete-nearly-complete}
\end{figure}

To assess the scalability of our implementation in the face of
particularly difficult UBA, we considered two families of
parametrized UBA. Both have an alphabet defined over a single atomic
proposition resulting in a two-element alphabet that we use to
represent either a $0$ (meaning the negated atomic proposition) or a $1$ bit
(meaning the positive atomic proposition). The first automaton
(``complete automaton''),
depicted in Figure~\ref{fig:uba-complete-nearly-complete} on the left for $k=2$,
is a complete automaton, i.e.,
recognizes $\Sigma^\omega$. It consists of a single,
accepting starting state that nondeterministically branches to one
of $2^k$ states, each one leading after a further step
to a \(k\)-state chain that only lets a particular $k$-bit bit-string
pass, subsequently returning to the initial state. As all the $k$-bit
bit-strings that can occur have a chain, the automaton is
complete. Likewise, the automaton is unambiguous as each of the
bit-strings can only pass via one of the chains.

Our second automaton (``nearly complete automaton''),
depicted in Figure~\ref{fig:uba-complete-nearly-complete} on the right for $k=2$,
arises from the first automaton by a modification of the
chain for the ``all zero'' bit-string, preventing the return to the
initial state. Clearly, the automaton is not complete.

We use both kinds of automata in an experiment using our extension of
\prism{} against a simple, two-state DTMC that encodes a uniform
distribution between the two ``bits''. This allows us to determine
whether the given automaton is almost universal. As the \prism{}
implementation requires the explicit specification of a DTMC, we end
up with a product that is slightly larger than the UBA, even though we
are essentially performing the UBA computations for the uniform
probability distribution.
In particular, this
experiment serves to investigate the scalability of our implementation
in practice for determining whether an SCC is positive, for the cut
generation and for computing $\vec{z}$.
It should be noted that equivalent deterministic automata, e.g.,
obtained by determinizing the UBA using the \ltltodstar{} tool, are significantly
smaller (in the range of tens of states) due to the fact that the UBA
in question are constructed inefficiently on purpose.

Table~\ref{tab:bench-complete-automaton} presents statistics for
our experiments with the ``complete automaton'' with various parameter
values $k$, resulting in increasing sizes of the UBA and the SCC
(number of states). We list the time spent for generating a cut
($t_\mathrm{cut}$), the number of iterations in the cut generation algorithm of
Lemma~\ref{lem:compute-cut}, and the size of the
cut. In all cases, the cut generation requires 2 iterations. Then we
compare the SCC handling based on power iteration with the SCC
handling relying on a rank computation for determining positivity of
the SCC and a subsequent computation of the values. For the power
iteration method, we provide the time spent for iteratively computing
an eigenvector ($t_\mathrm{eigen}$) and the number of iterations
(iter.).
For the other method, we provide the time spent for the positivity
check by a rank computation with a QR decomposition from the \colt{}
library ($t_\mathrm{positive}$) and for the subsequent computation of
the values via solving the linear equation system
($t_\mathrm{values}$). We used an overall timeout of 60 minutes for
each \prism{} invocation and an epsilon value of $10^{-10}$ as the
convergence threshold.

As can be seen, the power iteration method for the numeric SCC
handling performs well, with a modest increase in the number of
iterations for rising~$k$ until converging on an
eigenvector, as it can fully exploit the sparseness of the matrix.
The QR decomposition for rank computation
performs worse. The time for cut generation exhibits a super-linear
relation with $k$, which is reflected in the larger number of words
that were checked to determine that they are an extension. Note that
our example was chosen in particular to put stress on the cut
generation.

The results for the ``nearly complete automaton'', shown in
Table~\ref{tab:bench-nearly-complete-automaton}, focus on the
computation in the ``dominant SCC'', i.e., the one containing all the
chains that return to the initial state. For the other SCC,
containing the self-loop, non-positivity is immediately clear as it
does not contain a final state. In contrast to the ``complete
automaton'', no cut generation takes place, as the SCC is not
positive. The results roughly mirror the ones for the ``complete
automata'', i.e., the power iteration method is quite efficient in
determining that the SCC is not positive, while the QR decomposition
for the rank computation needs significantly more time and scales
worse.

As the power iteration method performed better, our benchmark results
presented in the following subsections use this method for the SCC handling.

\begin{landscape}
\begin{table}[tbp]
\centering
\begin{tabular}{r|r|r|r|r|r|r|r|r|r}
  &
  &
  & \multicolumn{3}{c|}{cut generation}
  & \multicolumn{2}{c|}{power iter.}
  & \multicolumn{2}{c}{rank-based}
  \\
   $k$
 & \,\(\vert\cA\vert\)
 & \,SCC size
 & $t_\mathrm{cut}$
 & \,ext.\ checks
 & \,cut size
 & \,$t_\mathrm{eigen}$
 & iter.
 & \,$t_\mathrm{positive}$
 & \,$t_\mathrm{values} $
\\ \hline
%
   5
 & 193
 & 258
 & \psec{0.061}
 & 10124
 & 32
 & \subpointone 
 & 215
 & \psec{0.451}
 & \psec{0.368}
 \\

   6
 & 449
 & 578
 & \psec{0.123}
 & 40717
 & 64
 & \subpointone 
 & 282
 & \psec{4.329}
 & \psec{4.329}
 \\

   7
 & 1025
 & 1282
 & \psec{0.876}
 & 172102
 & 128
 & \psec{0.07}
 & 358
 & \psec{56.517}
 & \psec{56.862}
 \\

   8
 & 2305
 & 2818
 & \psec{1.818}
 & 929413
 & 256
 & \psec{0.088}
 & 443
 & \psec{830.796}
 & \psec{835.121}
 \\

   9
 & 5121
 & 6146
 & \,\psec{17.919}
 & 6818124
 & 512
 & \psec{0.142}
 & 537
 & -\ \
 & -\ \
 \\
\end{tabular}
\caption{Benchmark results for ``complete automaton''
         with parameter $k$. \(-\) stands for time-out.}
\label{tab:bench-complete-automaton}
\end{table}

\begin{table}[btp]
\centering
\begin{tabular}{r|r|r|r|r|r}
  &
  &
  & \multicolumn{2}{c|}{power iteration}
  & \multicolumn{1}{c}{rank-based}
  \\
   $k$
 & \ \(\vert\cA\vert\)
 & \ SCC size
 & \ $t_\mathrm{eigen}$
 & \ iter.
 & \ $t_\mathrm{positive}$
\\ \hline
%
   5
 & 193
 & 250
 & \subpointone 
 & 52
 & \psec{0.399}
 \\

   6
 & 449
 & 569
 & \subpointone 
 & 78
 & \psec{4.105}
 \\

   7
 & 1025
 & 1272
 & \subpointone 
 & 112
 & \psec{54.435}
 \\

   8
 & 2305
 & 2807
 & \psec{0.072}
 & 155
 & \psec{844.016}
 \\

   9
 & 5121
 & 6134
 & \psec{0.112}
 & 205
 & -\ \
 \\
\end{tabular}
\caption{Benchmark results for ``nearly complete automaton''
         with parameter $k$}
\label{tab:bench-nearly-complete-automaton}
\end{table}

\end{landscape}




\subsection{Case Study: Bounded Retransmission Protocol}

Next we report on benchmarks using the bounded retransmission protocol (BRP)
case study of the \prism{} benchmark suite~\cite{prismBenchmark}.  The model
from the benchmark suite covers a single message transmission, retrying for a
bounded number of times in case of an error. In this protocol the message is
split into several so-called hunks. The number of hunks and the number of
allowed retransmissions are parameters in the model. We set the number of hunks
to 
$16$ and the maximal of retransmissions to $128$.

We have slightly modified the model
to allow the transmission of an infinite number of messages by restarting the
protocol once a message has been successfully delivered or the bound for
retransmissions has been reached.  We include benchmarks with pre-generated
automata, as well as benchmarks with LTL as starting point.
We include also the
evaluation for deterministic Rabin automata generated by \rabinizer{} from \cite{KMSZ18}.

\paragraph{Automata based specifications}
We consider the property ``the message was retransmitted \(k\) steps before an
acknowledgment.''
To remove the effect of selecting specific tools for the LTL to automaton
translation (\ltltotgba{} for UBA, the Java-based \prism{}
reimplementation of \ltltodstar{}~\cite{KB06} to obtain a
deterministic Rabin automaton (DRA) for the \prism{} standard approach),
we first consider model checking directly against automata specifications.
As the language of the property is equivalent to the UBA depicted in
Figure~\ref{fig:uba-examples} (on the left) where
\(a\) stands for a retransmission, \(b\) for an acknowledgment, and \(c\) for
no acknowledgment, we use this automaton and the minimal DBA for
the language (this case is denoted by $\cA^k$). We additionally consider
the UBA and DBA obtained by replacing the self-loop in the last state with a
switch back to the initial state (denoted by $\cB^k$), i.e., roughly applying
the $\omega$-operator to $\cA^k$.



\begin{landscape}
\begin{table}[t]
    \centering
\begin{tabular}{r||r|r|r||r|r|r|r}

   &
   \multicolumn{3}{c||}{\prism{} standard} &
   \multicolumn{4}{c}{\prism{} UBA}\\

   &
   \(\vert\cA^k_\textit{DRA}\vert\) &
   \(\vert\cM \otimes \cA^k_\textit{DRA}\vert\) &
   $t_{MC}$ &
   \(\vert\cA^k_\textit{UBA}\vert\) &
   \(\vert\cM \otimes \cA^k_\textit{UBA}\vert\) &
   $t_{\mathit{MC}}$ &
   $t_{\mathit{Pos}}$\\\hline
\(k=4\), \(\mathcal{A}^4\) &
\pnodes{33} &
\pnodes{61025} &
\psec{0.442} &
\pnodes{6} &
\pnodes{34118} &
\psec{0.251} &
\\
\(\mathcal{B}^4\) &
\pnodes{33} &
\pnodes{75026} &
\psec{0.398} &
\pnodes{6} &
\pnodes{68474} &
\psec{1.348} &
\psec{1.022}
\\\hline
\(k=6\), \(\mathcal{A}^6\) &
\pnodes{129} &
\pnodes{62428} &
\psec{0.481} &
\pnodes{8} &
\pnodes{36164} &
\psec{0.249} &
\\
\(\mathcal{B}^6\) &
\pnodes{129} &
\pnodes{97754} &
\psec{0.499} &
\pnodes{8} &
\pnodes{99460} &
\psec{1.71} &
\psec{1.325}
\\\hline
\(k=8\), \(\mathcal{A}^8\) &
\pnodes{513} &
\pnodes{64715} &
\psec{0.619} &
\pnodes{10} &
\pnodes{38207} &
\psec{0.261} &
\\
\(\mathcal{B}^8\) &
\pnodes{513} &
\pnodes{134943} &
\psec{0.713} &
\pnodes{10} &
\pnodes{136427} &
\psec{2.595} &
\psec{2.11}
\\\hline
\(k=14\), \(\mathcal{A}^{14}\) &
\pnodes{32769} &
\pnodes{83845} &
\psec{4.162} &
\pnodes{16} &
\pnodes{44340} &
\psec{0.31} &
\\
\(\mathcal{B}^{14}\) &
\pnodes{32769} &
\pnodes{444653} &
\psec{4.879} &
\pnodes{16} &
\pnodes{246346} &
\psec{6.817} &
\psec{6.078}
\\\hline
\(k=16\), \(\mathcal{A}^{16}\) &
\pnodes{131073} &
\(-\) &
\(-\) &
\pnodes{18} &
\pnodes{46390} &
\psec{0.322} &
\\
\(\mathcal{B}^{16}\) &
\pnodes{131037} &
\(-\) &
\(-\) &
\pnodes{18} &
\pnodes{282699} &
\psec{8.885} &
\psec{7.96}
\\\hline
\(k=48\), \(\mathcal{A}^{48}\) &
\(-\) &
\(-\) &
\(-\) &
\pnodes{50} &
\pnodes{79206} &
\psec{0.825} &
\\
\(\mathcal{B}^{48}\) &
\(-\) &
\(-\) &
\(-\) &
\pnodes{50} &
\pnodes{843414} &
\psec{72.432} &
\psec{70.286}
\end{tabular}
    \caption{Statistics for DBA/DRA- and UBA-based model checking of
      the BRP case study, a
      DTMC with $29358$ states,
      showing the number of
      states for the automata and the product
      and the time for model checking ($t_\textit{MC}$).
      For $\cB$, the time for checking positivity ($t_{\mathit{Pos}}$)
      is included in $t_\textit{MC}$.
      The mark $-$ stands for ``not available'' or timeout (30 minutes).}
    \label{table:brp-aut}
\end{table}
\end{landscape}

Table \ref{table:brp-aut} shows results for selected $k$ (with a timeout of $30$ minutes),
demonstrating that for this case study and properties
our UBA-based implementation is generally competitive with the
standard approach of \prism{} based on deterministic automata.
For $\cA^k$, our implementation detects that the UBA has a
special shape where all final states have a true-self loop which
allows skipping the SCC handling.
If we execute the positivity check nevertheless, $t_{\mathit{Pos}}$ is in the sub-second range for all considered $\cA^k$.
At a certain point, the implementation of the standard approach in
\prism{} becomes unsuccessful, due to \prism{} size
limitations in the product construction of the Markov chain and the
deterministic automaton
($\cA^k$/$\cB^k$: $k\geqslant 16$).
As can be seen, using the UBA approach
we can scale the parameter $k$ beyond $48$
when dealing directly with the automata-based specifications
($\cA^k$/$\cB^k$) and within reasonable time required for model checking.

\paragraph{LTL based specifications}

We consider two LTL properties: The first one is:
\[\varphi^k = (\neg \texttt{ack\_received}) \ \until \ \bigl(\texttt{retransmit} \wedge (\neg \texttt{ack\_received} \ \until^{= k}\ \texttt{ack\_received})\bigr),\]
where \(a \until^{=k} b\) stands for \(a \wedge \neg b \wedge \neXt (a \wedge
\neg b) \wedge \ldots \wedge \neXt^{k-1} (a \wedge \neg b) \wedge \neXt^k b\).
The formula~\(\varphi^k\)
ensures that $k$ steps before an acknowledgment the message was retransmitted.
Hence, it is equivalent to the property described by the automaton~\(\cA^k\).
For the LTL-to-automaton translation we
included the Java-based \prism{} reimplementation of \ltltodstar{}
\cite{KB06} to obtain a deterministic Rabin automaton (DRA) for the \prism{} standard approach as well as the tool \rabinizer{} (version 3.1) from \cite{EsparzaKS16}. For the
generation of UBA, we relied on \spot{} (version 2.7), as it is
the only tool that is capable of generating UBA explicitly. 
\begin{landscape}
\begin{table}[tbp]
\centering
\begin{tabular}{r||r|r|r||r|r|r||r|r|r}
   k
   &
   \multicolumn{3}{c||}{\prism{} standard} &
   \multicolumn{3}{c||}{\prism{} \rabinizer} &
   \multicolumn{3}{c}{\prism{} UBA}\\

   &
   \(\cA_\textit{DRA}\) &
   \(\vert \cM \otimes \cA_\textit{DRA}\vert\) &
   $t_{MC}$ &
   \(\cA_\textit{Rab}\) &
   \(\vert \cM \otimes \cA_\textit{Rab}\vert\) & 
   $t_{MC}$ &
   \(\cA_\textit{UBA}\) &
   \(\vert \cM \otimes \cA_\textit{UBA}\vert\) & 
   $t_{\mathit{MC}}$
   \\\hline
\(4\) & 
\pnodes{122} & 
\pnodes{62162} & 
\psec{1.678} & 
\pnodes{18} & 
\pnodes{60642} & 
\psec{0.568} & 
\pnodes{6} & 
\pnodes{34118} & 
\psec{0.628} 
\\\hline
\(6\) & 
\pnodes{4602} & 
\pnodes{72313} & 
\psec{3.336} & 
\pnodes{66} & 
\pnodes{61790} & 
\psec{0.641} & 
\pnodes{8} & 
\pnodes{36164} & 
\psec{0.534} 
\\\hline
\(8\) & 
\(-\) & 
\(-\) & 
\(-\) & 
\pnodes{258} & 
\pnodes{63698} & 
\psec{1.049} & 
\pnodes{10} & 
\pnodes{38207} & 
\psec{0.568} 
\\\hline
\(10\) & 
\(-\) & 
\(-\) & 
\(-\) & 
\pnodes{1026} & 
\pnodes{66739} & 
\psec{3.754} & 
\pnodes{12} & 
\pnodes{40249} & 
\psec{0.672} 
\\\hline
\(12\) & 
\(-\) & 
\(-\) & 
\(-\) & 
\pnodes{4098} & 
\pnodes{71660} & 
\psec{38.514} & 
\pnodes{14} & 
\pnodes{42293} & 
\psec{1.006} 
\\\hline
\(14\) & 
\(-\) & 
\(-\) & 
\(-\) & 
\pnodes{16386} & 
\pnodes{79576} & 
\psec{925.455} & 
\pnodes{16} & 
\pnodes{44340} & 
\psec{5.837} 
\\\hline
\(16\) & 
\(-\) & 
\(-\) & 
\(-\) & 
\(-\) & 
\(-\) & 
\(-\) & 
\pnodes{18} & 
\pnodes{46390} & 
\psec{132.873} 
\end{tabular}
\caption{Statistics for automata-based (standard, \rabinizer{}, and UBA)
    model checking of the BRP model and \(\varphi^k\). For every approach the
    corresponding automata sizes and product sizes are depicted.
     The overall model
    checking times (\(t_\textit{MC}\)) are listed, which includes the time for
    automata translation.}
\label{table:brp-ltl1}
\end{table}

\begin{table}[btp]
\centering
\begin{tabular}{r||r|r|r||r|r|r||r|r|r|r|r}
   k
   &
   \multicolumn{3}{c||}{\prism{} standard} &
   \multicolumn{3}{c||}{\prism{} \rabinizer} &
   \multicolumn{5}{c}{\prism{} UBA}\\

   &
   \(\cA_\textit{DRA}\) &
   \(\vert \cM \otimes \cA_\textit{DRA}\vert\) &
   $t_{MC}$ &
   \(\cA_\textit{Rab}\) &
   \(\vert \cM \otimes \cA_\textit{Rab}\vert\) & 
   $t_{MC}$ &
   \(\cA_\textit{UBA}\) &
   \(\vert \cM \otimes \cA_\textit{UBA}\vert\) & 
   \(t_{pos}\) &
   \(t_{Cut}\) &
   $t_{\mathit{MC}}$
   \\\hline
\(1\) & 
\pnodes{6} & 
\pnodes{29358} & 
\psec{0.682} & 
\pnodes{5} & 
\pnodes{29358} & 
\psec{0.435} & 
\pnodes{4} & 
\pnodes{31422} & 
\subpointone & 
n/a & 
\psec{0.315} 
\\\hline
\(2\) & 
\pnodes{17} & 
\pnodes{37678} & 
\psec{0.945} & 
\pnodes{7} & 
\pnodes{35630} & 
\psec{0.507} & 
\pnodes{8} & 
\pnodes{41822} & 
\psec{4.84} & 
\psec{0.192} & 
\psec{5.435} 
\\\hline
\(3\) & 
\pnodes{65} & 
\pnodes{39726} & 
\psec{1.117} & 
\pnodes{11} & 
\pnodes{37678} & 
\psec{0.537} & 
\pnodes{14} & 
\pnodes{45934} & 
\psec{5.151} & 
\psec{0.204} & 
\psec{5.78} 
\\\hline
\(4\) & 
\pnodes{314} & 
\pnodes{43806} & 
\psec{1.523} & 
\pnodes{23} & 
\pnodes{41758} & 
\psec{0.624} & 
\pnodes{22} & 
\pnodes{54126} & 
\psec{5.785} & 
\psec{0.287} & 
\psec{6.565} 
\\\hline
\(5\) & 
\pnodes{1443} & 
\pnodes{47902} & 
\psec{2.329} & 
\pnodes{59} & 
\pnodes{45854} & 
\psec{0.894} & 
\pnodes{32} & 
\pnodes{62334} & 
\psec{6.523} & 
\psec{0.255} & 
\psec{7.29} 
\\\hline
\(6\) & 
\pnodes{9016} & 
\pnodes{56029} & 
\psec{5.273} & 
\pnodes{167} & 
\pnodes{53997} & 
\psec{2.097} & 
\pnodes{44} & 
\pnodes{78669} & 
\psec{9.484} & 
\psec{0.214} & 
\psec{10.294} 
\\\hline
\(7\) & 
\pnodes{67964} & 
\(-\) & 
\(-\) & 
\pnodes{491} & 
\pnodes{58081} & 
\psec{9.579} & 
\pnodes{58} & 
\pnodes{86853} & 
\psec{10.351} & 
\psec{0.214} & 
\psec{11.307} 
\\\hline
\(8\) & 
\(-\) & 
\(-\) & 
\(-\) & 
\pnodes{1463} & 
\pnodes{66217} & 
\psec{76.115} & 
\pnodes{74} & 
\pnodes{103157} & 
\psec{13.852} & 
\psec{0.266} & 
\psec{14.984} 
\\\hline
\(9\) & 
\(-\) & 
\(-\) & 
\(-\) & 
\pnodes{4379} & 
\pnodes{70291} & 
\psec{783.666} & 
\pnodes{92} & 
\pnodes{111321} & 
\psec{15.23} & 
\psec{0.284} & 
\psec{16.786} 
\\\hline
\(10\) & 
\(-\) & 
\(-\) & 
\(-\) & 
\(-\) & 
\(-\) & 
\(-\) & 
\pnodes{112} & 
\pnodes{127562} & 
\psec{20.095} & 
\psec{0.301} & 
\psec{22.707} 
\end{tabular}%
\caption{Statistics for automata-based (standard, \rabinizer{}, and UBA) of the BRP
model and \(\psi^k\). The structure of this table corresponds to
Table~\ref{table:brp-ltl1}, but with additional listing of the time for the
positivity checks \(t_\mathit{pos}\) and cut calculation time
\(t_\mathit{cut}\). n/a means not available.}
\label{table:brp-ltl2}
\end{table}
\end{landscape}

Table~\ref{table:brp-ltl1} lists the results for model checking \(\varphi^k\).
From a certain point on, the implementation of the standard approach in \prism{}
is unsuccessful, due to \prism's restrictions in the DRA construction
($k\geqslant 8$). Concerning automata sizes and model checking times, \spot{}
shows the best behavior among \prism{} standard and \rabinizer{}.  \spot{}
actually generates a UFA for \(\varphi^k\) which is recognized by our
implementation and handled as explained in \cite{BenLenWor14}.  The sizes of the
unambiguous automata output by \spot{} grow linearly in $k$, whereas the
sizes of the deterministic automata output by \rabinizer{} grow
exponentially. Thus, \prism{} with \rabinizer{} times out after $30$
minutes for $k=15$.  However, \spot{} produces for $\varphi^k$ an
exponential-sized intermediate automaton, which is then shrunk via
bisimulation to an automaton of linear size. Thus, our implementation \prism{}
UBA times out for $k=18$.


As a second formula we investigate

\[\psi^k = \globally (\texttt{msg\_send} \rightarrow \finally
(\texttt{ack\_send} \wedge \finally^{\leqslant k} \texttt{ack\_received})),\]

where \(\finally^{\leqslant k} a\) denotes \(a \vee \underbrace{\neXt (a \vee
\neXt ( \ldots \vee \neXt a))}_{k \text{ times}}\). This formula requires that every request (sending a message and waiting for an
acknowledgment) is eventually responded to by an answer (the receiver of the
message sends an acknowledgment and this acknowledgment is received within the
next \(k\) steps).

Table~\ref{table:brp-ltl2} summarizes the results of the benchmark for
\(\psi^k\).  Here, the \prism{} standard approach with its own implementation of
\ltltodstar{} finishes the calculations until \(k=6\). The sizes of
the DRA produced by {\prism}'s \ltltodstar{} increase rapidly with~\(k\).
For \(k=7\), \prism{} standard can construct the DRA (with
\pnodes{67964} states and within \(37.0\) seconds), but cannot construct
the product anymore.  Similarly, the sizes of the DRA produced by \rabinizer{}
grow rapidly in \(k\). The sizes of the DRA of \rabinizer{} are smaller than the
size of the UBA for $k \geqslant 4$.

In contrast to the deterministic automata, the UBA sizes increase moderately with~\(k\).
In the UBA approach the positivity check is the most time
consuming part of the calculation, whereas the cut generation is always below
\(0.4\) seconds. For \(k=1\) there is no positive SCC, so the
cut calculation is omitted. The model checking process consumes more time
in the UBA case in comparison with \prism{} standard until \(k=6\), but for
bigger \(k\) the performance turns around. Even if \prism{} standard were to
complete the calculation for \(k=7\), it would be slower, as the creation
of the DRA takes \(37.0\) seconds. Similarly, \prism{} \rabinizer{} outperforms
\prism{} UBA for $k \leqslant 8$, which is due to the time-consuming positivity
check. For bigger \(k\), both \prism{} standard and \prism{} \rabinizer{} cannot
finish their calculation within the given time bound, whereas \prism{} UBA
finishes the calculations for all tested $k \leqslant 12$. 

\subsection{NBA versus UBA}
\label{sec:nbavsuba}

To gain some understanding of the cost of requiring unambiguity for an
NBA, we compare the sizes of NBA and UBA generated by the
\ltltotgba{} tool of \spot{} for the formulas
of~\cite{EH00,SomBloem00,DwyerAC99}, which have been used for benchmarking, e.g.,
in~\cite{KB06}. We consider both the ``normal'' formulas and their
negations, yielding 188 formulas.

\begin{table}[htbp]
     \centering
     \resizebox{\textwidth}{!}{%
     \begin{tabular}{@{}r|r|r|r|r|r|r|r|r|r}
        Number of states $\leqslant x$& $\leqslant 1$& $\leqslant 2$& $\leqslant 3$& $\leqslant 4$& $\leqslant 5$& $\leqslant 7$& $\leqslant 10$& $\leqslant 20$& $\geqslant 20$\\\hline
        \ltltotgba{} NBA & $12$ & $49$ & $103$ & $145$ & $158$ & $176$ & $181$ & $188$ & $0$\\
        \ltltotgba{} UBA & $12$ & $42$ & $74$ & $108$ & $123$ & $153$ & $168$ & $180$ & $8$
     \end{tabular}%
     }
     \caption{Number of formulas where the (standard) NBA and UBA has a number
     of states $\leqslant x$.}
     \label{table:nbavsuba}
 \end{table}

As can be seen in Table~\ref{table:nbavsuba}, both the NBA and UBA are of reasonable size.
Most of the generated UBA ($102$) have the same size as the NBA and
for $166$ of the formulas the UBA is at most twice the size
as the corresponding NBA. The largest UBA has 112 states, the second
largest has 45 states.





\section{Conclusion}
\label{sec:conclusion}

We have presented a polynomial-time algorithm for Markov chain
analysis against properties given as unambiguous automata.  The
algorithm is based on the analysis of nonnegative matrices and
exploits in particular their spectral theory.

As LTL formulas can be transformed into UBA with a single exponential
blow-up, our algorithm yields a procedure for model checking LTL
formulas on Markov chains that is singly exponential in the formula
size.  We have moreover refined the process of UBA generation and
Markov chain analysis to achieve the optimal PSPACE upper bound for
computing the exact probability that a given Markov chain satisfies a
given LTL formula.


We have developed an extension of \prism{} that supports our approach.
Its experimental evaluation shows that the UBA-based method is very
competitive with the approach using deterministic automata,
outperforming the latter in certain cases.

For the other singly exponential approaches to LTL model checking,
such as using separated automata~\cite{CouSahSut03} or weak
alternating automata~\cite{BusRubVar04}, we are not aware of any
available implementation to compare our approach against.%
\footnote{The paper \cite{CouSahSut03} addresses experiments with a
  prototype implementation, but this implementation seems not to be
  available anymore.  } Our algorithm for arbitrary UBA can be seen as
a generalization of the approach of~\cite{CouSahSut03}, which requires
separated UBA~\cite{MuellerThesis18}.

Markov chain analysis via general UBA, rather than separated UBA or
deterministic automata, offers additional flexibility that can be
exploited when building automata from LTL formulas.  In particular, it
facilitates use of state-reduction techniques, such as simulation,
that may not preserve the seperatedness property.  As our experiments
(specifically with the bounded retransmission protocol) suggest, the
eigenvalue algorithm can deduce non-positiveness of an SCC very
efficiently in practice.  For the generation of UBA we have used the
tool \spot, which implements a simple and straightforward way to
produce unambiguous B\"uchi automata \cite{DuretHabil17}.
Alternatively, \texttt{Tulip} contains an LTL-to-UBA translator; but
this tool is not available anymore.  As with the approach using
deterministic automata, the performance of the UBA-based method
depends strongly on the availability of small UBA. In contrast to
nondeterministic or deterministic automata, the generation of small
UBA and their simplification has not yet been explored thoroughly.

\bibliographystyle{plain}
\bibliography{lit}

\end{document}